\documentstyle[prb,aps,12pt]{revtex}

\begin{document}
\title{Spectral properties and infrared absorption in manganites}
\author{C.A. Perroni, G. De Filippis, V. Cataudella and G. Iadonisi}
\address{INFM  and Dipartimento di Scienze Fisiche, \\
Universit\`{a} degli Studi di Napoli ``Federico II'',\\
Complesso Universitario Monte Sant'Angelo,\\
Via Cintia, I-80126 Napoli, Italy}
\date{\today}
\maketitle

\begin{abstract}
Within a recently proposed variational approach it has been shown that, in 
$La_{1-x}A_xMnO_3$ perovskites with $0<x<0.5$, near the metal-insulator 
transition, the combined effect of the magnetic and electron-phonon 
interactions pushes the system toward a regime of two coexisting phases: a low
electron density one made by itinerant large polarons forming ferromagnetic 
domains and a high electron density one made by localized small polarons 
giving rise to paramagnetic or antiferromagnetic domains depending on 
temperature.
Employing the above-mentioned variational scheme, in this paper spectral and 
optical properties of manganites are derived for $x=0.3$ at different 
temperatures.
It is found that the phase separation regime induces a robust pseudogap in the 
excitation spectrum of the system.
Then the conductivity spectra are characterized by a transfer of spectral 
weight from high to low energies, as the temperature $T$ decreases.    
In the metallic ferromagnetic phase, at low $T$ two types of infrared 
absorption come out: a Drude term and a broad absorption band due respectively
to the coherent and incoherent motion of large polarons.
The obtained results turn out in good agreement with experiments.
\end{abstract}


\newpage 
In the last years the perovskite oxides $La_{1-x}A_xMnO_3$ ($A$ represents
a divalent alkali element such as $Sr$ or $Ca$)
have become one of the main research areas of the condensed matter
community because of the colossal magnetoresistance effect exhibited for
$0.2 \leq x \leq 0.5$.
\cite{jin,dagotto}
The study of these materials started in the 1950's showing the strong 
correlation between magnetization and resistivity. \cite{1} 
Since in the electronically active Mn $3d$ orbitals the mean number of $d$
electrons per $Mn$ is $4-x$, three electrons go into the $t_{2g}$ core states
and the remaining $1-x$ electrons occupy the outer shell $e_g$ orbitals.
The ferromagnetic phase was explained by introducing the double exchange 
mechanism\cite{2,2bis} that takes into account the combined effect of the 
$e_g$ electron hopping between nearest neighbor sites and the very strong
Hund's exchange with the localized $t_{2g}$ electron spins.    
In the 1990's the discovery of the colossal magnetoresistance phenomenon 
has aroused a renewed interest for these compounds. 
In order to explain the experimental data, it has been suggested that, 
in addition to the double-exchange term, a strong interaction  
between electrons and lattice distortions plays a non negligible role.\cite{3} 
Next the relevance of the Jahn-Teller polaron formation has been confirmed
experimentally by the 
giant isotope shift of the Curie temperature,\cite{5} by measurements of the
lattice distortions by EXAFS\cite{6} and by frequency shifts of the internal
phonon modes.\cite{7}
Furthermore both pseudogap features and conductivity spectra of these 
compounds have been discussed in terms of a strong coupling to lattice 
distortions.\cite{dessau1,dessau2,kim1,kim2,yoon,machida} 
Nowadays the role of the polaron is widely recognized.\cite{millis}

Recent studies have showed the presence of strong tendencies toward the phase 
separation in manganites.\cite{dagotto,9}
At low temperatures for compositions near the metal-insulator transition 
the ferromagnetic and antiferromagnetic phases coexist while, 
close to the paramagnetic to ferromagnetic transition temperature, 
the phase separation occurs between paramagnetic and ferromagnetic domains. 
From a theoretical point of view, the coexistence between hole-poor
and hole-rich phases has been discussed by using exact numerical approaches 
on small lattices assuming classical Jahn-Teller phonons.\cite{dagotto,moreo} 
The pseudogap formation\cite{moreo1} and the optical properties
\cite{moreo2,millis2} also have been studied neglecting the quantum nature of 
phonons. 
One of the drawbacks of this approximation in the optical properties is, for
example, that at low temperatures in the metallic ferromagnetic phase a 
narrow Drude peak cannot be obtained.
   
In a recent paper\cite{cata}, some of us have shown, on the basis of a 
variational approach, that the quantum character of the Jahn-Teller phonons 
and the polaron formation can be important to explain the experimentally 
observed tendency of manganites to form inhomogeneous magnetic structures near 
the phase boundaries. We employ the scheme proposed in that paper to deduce 
spectral and optical properties of manganites for $x=0.3$ at different 
temperatures. It is interesting to study the properties of these compounds at 
this doping since in the range of concentrations around $x=0.3$ the colossal 
magnetoresistance effect is very pronounced in many manganese oxides.
\cite{jin} 
In this paper we find that the phase separation regime induces a robust 
pseudogap in the 
excitation spectrum of the system. The pseudogap features show a direct 
relationship with the metal-insulator transition and with the crossover from 
the coherent large polaron dynamics to the incoherent small polaron one. Our 
results turn out compatible with experimental findings.\cite{dessau1,dessau2}
Next, by using the formalism of generalized Matsubara Green's functions
\cite{schna,loos1,loos2,kada,fehske},
we determine the scattering rate of charges affected by the interaction
with the lattice and the spin fluctuations. 
The single phonon emission and absorption represent the main mechanism
of scattering. However, the damping due to spin fluctuations is effective in 
the energy range around the chemical potential $\mu$.
The scattering rate turns out fundamental to derive the very interesting 
optical properties of the system.
With decreasing $T$, our spectra are characterized by a 
transfer of spectral weight from high to low energies filling up the low 
frequency optical gap present in the high-temperature phase.
Indeed at high $T$ the infrared absorption is due to the incoherent
small polaron dynamics.    
Instead, at low temperatures, in the ferromagnetic phase the system shows two 
types of optical response: a Drude term and a broad absorption band due 
respectively to the coherent and incoherent motion of large polarons.
The results obtained are consistent with experimental data.
\cite{kim1,kim2,okim}

In section 1 the model and the variational approach are reviewed; in section
2 the spectral properties are deduced; in section 3 the damping of the 
particle motion is calculated; in section 4 the optical properties are 
discussed.
In appendix details of the calculations employed to derive the optical 
properties are reported.

\section{The model and the variational approach}

The model takes into account the double-exchange mechanism, the coupling of 
the $e_g$ electrons to lattice distortions and the super-exchange 
interaction between neighboring localized $t_{2g}$ electrons.
The coupling to longitudinal optical phonons arises from the Jahn-Teller
effect that splits the $e_g$ double degeneracy.\cite{zhang}
Adopting the single orbital approximation (reasonable for $x<0.5$), the 
Hamiltonian reads
 
\begin{eqnarray}
H=&&-t\sum_{<i,j>} 
\left(\frac{S^{i,j}_0+1/2}{2 S+1}\right) c^{\dagger}_{i}c_{j}
 +\omega_0 \sum_{i}a^{\dagger}_{i}a_{i}
+g \omega_0 \sum_{i} c^{\dagger}_{i}c_{i} \left( a_{i}+a^{\dagger}_{i} 
\right)
  \nonumber \\
&& + \epsilon \sum_{<i,j>} \vec{S}_{i} \cdot \vec{S}_{j} 
- \mu \sum_{i} c^{\dagger}_{i} c_{i} .  \label{1r}
\end{eqnarray}
Here  $t$ is the transfer integral between nearest neighbor ($nn$)
sites $<i,j>$ for electrons occupying $e_g$ orbitals, 
$S^{i,j}_0$ is the total spin of the subsystem consisting of 
two localized spins on $ nn $ sites and the conduction electron, 
$S$ is the spin of the $t_{2g}$ core states $\left( S= 3/2 \right)$, 
$c^{\dagger}_{i} \left( c_{i} \right)$ creates (destroys) an electron with 
spin parallel to the ionic spin at the i-th site. The first term of the 
Hamiltonian describes the double-exchange mechanism in the limit where the 
intra-atomic exchange integral $J$ is far larger than the transfer integral 
$t$. 
Furthermore in eq.(\ref{1r})  $\omega_0$ denotes the frequency of the local 
optical phonon mode, $ a^{\dagger}_{i} \left( a_{i} \right)$ is the creation 
(annihilation) phonon operator at the site i,
the dimensionless parameter $g$ indicates the strength of the
electron-phonon interaction in the Holstein model \cite{12}, 
$\epsilon$ represents the antiferromagnetic super-exchange coupling between 
two $nn$ $t_{2g}$ spins and
$\mu$ is the chemical potential.

The hopping of electrons is supposed to take place between the equivalent 
$nn$ sites of a simple cubic lattice separated by the distance 
$|n-n^{\prime}|=a$. 
The units are such that the Planck constant $\hbar=1$, the Boltzmann constant
$k_B$=1 and the lattice parameter $a$=1.

Following the recently proposed variational scheme\cite{cata}, 
we perform two successive canonical transformations to treat the 
electron-phonon interaction variationally. 

The first is the variational Lang-Firsov  unitary transformation 
\cite{14,sil}
\begin{equation}
U_{1}=\exp \left[- g \sum_{j}  
\left( f c^{\dagger}_{j} c_{j} +\Delta \right) 
\left( a_{j}-a^{\dagger}_{j} \right) 
\right]                                            
\label{2r}
\end{equation}
where $f$ and $\Delta$ are variational parameters.
The quantity  $f$ represents the strength of the coupling between an electron 
and the phonon displacement on the same site, hence it measures the degree of 
the polaronic effect. On the other hand, the site-independent parameter 
$ \Delta $ denotes a displacement field describing static lattice distortions 
that are not influenced by the instantaneous position of the electrons.

The second canonical transformation is Bogoliubov-type  
\cite{18} 
\begin{equation}
U_{2}=\exp \left[-\alpha \sum_{j} \left( a^{\dagger}_{j}
a^{\dagger}_{j} -a_{j}a_{j} \right)\right]          
\label{3r}
\end{equation}
where $\alpha$ is a variational parameter.
It introduces correlations between the emission of successive virtual 
phonons by the electrons and it is responsible of the phonon frequency 
renormalization.

The transformed Hamiltonian $\tilde{H}=  U_{2} U_{1} H U_{1}^{-1} U_{2}^{-1} $ 
becomes 

\begin{eqnarray}
\tilde{H}=&& 
\sum_{<i,j>} {\cal C}_{i,j} c^{\dagger}_{i}c_{j} 
+\bar{\omega}_0 \sum_{i}a^{\dagger}_{i}a_{i}+
N \omega_0 \sinh^2\left(2\alpha\right)
+N \omega_{0} g^{2} \Delta^{2}    \nonumber \\
&& 
+\omega_0 \sinh\left(2 \alpha\right) \cosh\left( 2 \alpha\right) 
\sum_{i} \left( a^{\dagger}_{i}a^{\dagger}_{i} +a_{i}a_{i} \right)
-\omega_{0} g e^{2 \alpha} \Delta
\sum_{i} \left( a_{i} +a^{\dagger}_{i} \right) \nonumber \\
&& 
+\epsilon\sum_{<i,j>} \vec{S}_{i} \cdot \vec{S}_{j}
+ \sum_{i} c^{\dagger}_{i}c_{i} \left( \tilde{{\cal C}}_{i}+\eta-\mu\right).  
\label{4r}
\end{eqnarray}
where the operator ${\cal C}_{i,j}$ reads

\begin{equation}
{\cal C}_{i,j}= -t \left( \frac{S^{i,j}_0+1/2}{2 S+1}\right) 
X^{\dagger}_i X_j   \label{5r}
\end{equation} 
and $X_i$ is the phononic operator 

\begin{equation}
X_i=\exp \left[ g f e^{-2 \alpha} \left( a_{i} - a^{\dagger}_{i} \right)
\right].
\label{6r}
\end{equation}
Furthermore in the Hamiltonian (\ref{4r}) we specify the renormalized phonon 
frequency $ \bar{\omega}_0=\omega_0 \cosh(4 \alpha)$, the number of  lattice  
sites $N$, the phonon operator $\tilde{{\cal C}}_{i}$

\begin{equation}
\tilde{{\cal C}}_{i} =   
\omega_{0} g 
\left( 1-f \right) e^{2 \alpha} \left( a_{i} +a^{\dagger}_{i} \right)
\label{7r}
\end{equation} 
and the quantity $\eta$

\begin{equation}
\eta = \omega_{0} g^{2} f \left( f-2 \right)+2 \omega_{0} g^{2} 
\left( f-1 \right) \Delta 
\label{8r}
\end{equation}
that measures the electronic band shift due to the electron-phonon interaction.

To obtain the free energy in a variational scheme, we introduce a test 
Hamiltonian characterized by electron, phonon and spin 
degrees of freedom non mutually interacting

\begin{eqnarray}
H_{test} &=&
-t_{eff} \sum_{<i,j>} c^{\dagger}_{i}c_{j} +
\bar{\omega}_0 \sum_{i}a^{\dagger}_{i}a_{i}+
N \omega_0  \sinh^2\left( 2 \alpha\right)   
+N \omega_{0} g^{2} \Delta^{2}   \nonumber \\
&& 
-g_{s} \mu_{B}\sum_{i} \vec{h}_{eff} \cdot \vec{S}_{i}
+\left( \eta -\mu \right) \sum_{i} c^{\dagger}_{i}c_{i}.  
\label{9r}
\end{eqnarray}
The quantity $t_{eff}$ denotes the effective transfer integral

\begin{equation}
t_{eff}= t \left\langle \left(\frac{S_0+1/2}{2S+1}\right)\right\rangle 
e^{-S_{T}}
\label{10r}
\end{equation}
where the symbol $<>$ indicates a thermal average and the quantity $S_{T}$ is

\begin{equation}
S_{T}=g^{2} f^{2} e^{-4 \alpha} \left( 2N_0+1 \right)
\end{equation}
with  $N_0$ the average number of phonons with frequency $\bar{\omega}_0 $.
In the test Hamiltonian (\ref{9r}) $g_s$ is the dimensionless electron spin 
factor ($g_s \simeq 2 $), $\mu_B$ is the Bohr magneton and $h_{eff}$ is the
effective molecular magnetic field in a cell containing two neighboring sites.
The magnetic field is induced by a partial orientation of ionic spins around 
this cell and is determined by the variational approach.\cite{19}
We notice that in eq.(\ref{10r}) the factor $e^{-S_T}$ controls the band 
renormalization due to the polaron formation.

To obtain the variational free energy of the system, we employ the 
Bogoliubov inequality
\begin{equation}
F \leq F_{test}+\langle \tilde{H}-H_{test} \rangle _{t}
\end{equation}
where $<>_t$ indicates a thermodynamic average made using the test 
Hamiltonian. This approach allows to treat the local spin dynamics in a 
variational mean field theory.\cite{19}  
The free energy per site becomes 
\begin{eqnarray}
\frac{F}{N}=&& 
f_{test}^{el} + T \log{\left(1-e^{-\beta \bar{\omega}_0}\right)}+ 
\omega_0  \sinh^2\left( 2 \alpha\right) 
+\omega_{0} g^{2} (1-f)^2 \rho^2 
- T\log{\nu_S}     \nonumber \\
&&
\pm \frac{\epsilon}{2} Z S^2 m^2_S +  T\lambda m_S 
\label{11r}
\end{eqnarray}
where the electron free energy reads 

\begin{equation}
f_{test}^{el} = 
\left( \frac{1}{N} \right) \sum_{\bf{k}} \xi_{\bf{k}} n_{\bf{k}}
+T \left( \frac{1}{N} \right) \sum_{\bf{k}} \left[ n_{\bf{k}} \log n_{\bf{k}}+
\left( 1- n_{\bf{k}} \right)\log\left( 1- n_{\bf{k}}\right) \right]
\end{equation}
with $n_{\bf{k}}=n_{F} \left( \xi_{\bf{k}} \right) $ the Fermi distribution 
function.
We get $ \xi_{\bf{k}} = \bar{\varepsilon}_{\bf{k}} -\mu $, where
$ \bar{\varepsilon}_{\bf{k}} = \varepsilon_{\bf{k}}+\eta $
is the renormalized electronic band and $\varepsilon_{\bf{k}}$ the band 
dispersion

\begin{equation}
\varepsilon_{\bf{k}}=-2 t_{eff}[cos(k_x)+ cos(k_y)+cos(k_z)].
\label{11ur}
\end{equation}
Furthermore in eq.(\ref{11r}) $\beta$ is 
the inverse of the temperature, $\nu_S$ is the partition function of the 
localized spins, 
the top and bottom sign of $\epsilon$  hold, respectively, for the 
ferromagnetic and antiferromagnetic solutions of the localized spins,  
$Z$ indicates the number of nearest neighbors, $m_S$ represents the 
magnetization of the localized spins and $\lambda$ is a dimensionless 
variational parameter proportional to the effective magnetic field.

In order to simplify the calculations, we consider a semicircular electronic 
density of states

\begin{equation}
g(\xi)=\left( \frac{2}{\pi W^2} \right) 
\theta (W-| \xi -\eta +\mu| )   
\sqrt{ W^2-(\xi -\eta +\mu)^2 } 
\label{12r}
\end{equation}
where $W=Z t_{eff}$ is the renormalized band half-width and $\theta (x)$ is 
the Heaviside function. Actually $ g(\xi) $ represents a simple 
approximate expression for the exact density of states and it is 
generally used for a 3-D lattice. \cite{econo,georges} 
Therefore the electron free energy becomes  

\begin{eqnarray}
f_{test}^{el} =&&
\int_{-\infty}^{\infty} d\xi g(\xi) \xi n_{F}(\xi)
\nonumber \\
&&
+T \int_{-\infty}^{\infty}  d\xi g(\xi) 
\left\{ 
n_{F}(\xi)\log n_{F}(\xi)    
+\left[ 1- n_{F}(\xi) \right] \log \left[ 1- n_{F}(\xi) \right] 
\right\}.
\label{13r}
\end{eqnarray}

As shown in ref.\cite{cata}, in the intermediate electron-phonon 
coupling regime, the free energy (\ref{11r}) gives rise to a region of 
coexisting phases characterized by different 
electron densities $\rho_1$ and $\rho_2$ but the same value of chemical 
potential $\mu$.
The quantities $\rho_1$ and $\rho_2$ depend on the temperature and correspond 
to homogeneous phases of the system constituted by large and small polarons,
respectively.
Hence, near the metal-insulator transition (see Fig.1a), the system segregates 
in ferromagnetic and antiferromagnetic or paramagnetic domains of large and 
small polarons. We observe that the quantity $x$ denotes the 
hole concentration, that is $x=1- \rho$, where $\rho$ is the electron
concentration.

In the region of coexistence, the fractions of volume, $V_1/V$ and $V_2/V$,
filled with density $\rho _1$ and $\rho _2$, are determined by the two
conditions: $V_1/V+V_2/V=1$ and $(V_1/V)\rho_1+(V_2/V)\rho_2=\rho $. 
In Fig.1b we report the fractions of volume  $\left( V_1/V \right)_{ferro}$ 
and $\left( V_2/V \right)_{para}$ for x=0.3 as a function of the temperature
(the dotted line in Fig.1a allows to recognize the different phases of the 
system at this doping).

In the regime of phases coexisting,\cite{nagaev1,nagaev2} it is possible to 
 calculate any property $B$ of the
system by means of the linear combination of the properties 
$B_1$ and $B_2$ characteristic of the single phases assuming the respective 
volume fractions  $\left( V_1/V \right)$ and $\left( V_2/V \right)$ as 
weights 

\begin{equation}
B=\left( \frac {V_1}{V} \right) B_1+\left( \frac {V_2}{V} \right) B_2.
\label{14r}
\end{equation}
Hence we assume that in this regime the properties of the system are 
independent of morphology of coexisting domains.

\section{Spectral properties}

In this section we calculate the spectral properties of the system.

Performing the two canonical transformations (\ref{2r},\ref{3r}) and 
exploiting the cyclic properties of the trace, the electron Matsubara Green's 
function becomes \cite{mahan}

\begin{equation}
{\mathcal G} \left( \bf {k},\tau \right)=
- \left( \frac{1}{N} \right) \sum_{i,j} 
e^{ i \bf{k} \cdot \left( \bf{R}_i - \bf{R}_j \right) }  
\langle T_{\tau} \tilde{c}_i \left( \tau \right) \tilde{X}_i (\tau) 
c_j^{\dagger} X_j ^{\dagger}   \rangle
\label{16r}
\end{equation}
where 
$\tilde{c}_i \left( \tau \right)= e^{\tau \tilde{H}} c_i e^{-\tau \tilde{H}}$
and 
$\tilde{X}_i \left( \tau \right)= e^{\tau \tilde{H}} X_i e^{-\tau \tilde{H}}$,
with $\tilde{H}$ given by eq.(\ref{4r}).

The correlation function in (\ref{16r}) can be disentangled into electronic 
and phononic terms by using the test Hamiltonian (\ref{9r}), hence    

\begin{equation}
{\mathcal G} \left( \bf {k},\tau \right)=
- \left( \frac{1}{N} \right) \sum_{i,j} 
e^{ i \bf{k} \cdot \left( \bf{R}_i - \bf{R}_j \right) }  
\langle T_{\tau} \bar{c}_i \left( \tau \right) c_j^{\dagger} \rangle_t 
 \langle T_{\tau} \bar{X}_i \left( \tau \right) X_j^{\dagger} \rangle_t 
\label{16ar}
\end{equation}
where now we have 
$\bar{c}_i \left( \tau \right)= e^{\tau H_t} c_i e^{-\tau H_t}$ and
$\bar{X}_i \left( \tau \right)= e^{\tau H_t} X_i e^{-\tau H_t}$.

The Green's function in Matsubara frequencies $\omega_n$ reads

\begin{eqnarray}
{ \mathcal G} \left( {\bf k},i \omega_n \right)= &&  
 e^{-S_T} { \mathcal G}^{(0)} \left( {\bf k},i \omega_n \right)+ 
\nonumber \\
&& e^{ -S_T } \left( \frac{1}{N} \right)   \sum_{\bf{k}_1}
\left( \frac{1}{\beta} \right) \sum_{m} 
{\mathcal G}^{(0)} \left( {\bf k}_1,i \omega_m \right)
\int_0^{\beta} e^{(i \omega_n -i \omega_m ) \tau}
\{ e^{s\cosh\left[\bar{\omega}_0\left(\tau-\frac{\beta}{2}\right)\right]}
-1 \}
\label{17r}
\end{eqnarray}
where ${ \mathcal G}^{(0)} \left( {\bf k},i \omega_n \right)$ is the Green's 
function of non interacting particles and $s$ is
\begin{equation}
s= 2 f^2 g^2 e^{-4 \alpha} \left[ N_0 \left( N_0+1 \right) \right]
^{ \frac{1}{2} }.   
\label{19r}
\end{equation}
We notice that two physically distinct terms appear in the Eq.(\ref{17r}), the
coherent and the incoherent one.\cite{alex,ranni}
The first derives from the coherent motion of electrons and their 
surrounding phonon cloud (this process is called {\em diagonal transition}). 
The renormalization factor $e^{-S_T} $ appearing in this term 
plays the role of the residue $Z$ at the pole in a Fermi-liquid description.
\cite{imada}
The second contribution describes the possibility of changing  the
number of phonons in the phonon cloud during the electron hopping 
(this process is known as {\em non diagonal transition}). This is confirmed by
the presence of the sum over all electron momenta which reflects the fact that
the electron momentum is not conserved separately from the phonon one.

In the second term of the expression (\ref{17r}), we can perform the expansion

\begin{equation}
\exp \left\{
s\cosh\left[\bar{\omega}_0 \left(\tau-\frac{\beta}{2}\right)\right]
\right\}=
2 \sum_{l=1}^{+ \infty} I_l (s)
\cosh\left[\bar{\omega}_0\left(\tau-\frac{\beta}{2}\right)\right]+
I_0 (s)   
\label{20r} 
\end{equation}
into multiphonon contributions, where the symbols $ I_l (s) $ denote the 
modified Bessel functions. 

Making the analytic continuation $ i \omega_n \rightarrow \omega +i \delta $, 
one obtains the retarted Green's function $ G_{ret} ( {\bf k}, \omega)$ and 
the spectral function 

\begin{eqnarray}
A( {\bf k}, \omega) =&& -2 \Im G_{ret} ( {\bf k}, \omega) =
2 \pi e^{-S_T} \delta \left( \omega - \xi_{{\bf k}} \right)+
\nonumber \\
&& 2 \pi e^{-S_T}  \sum_{l(\neq 0)- \infty}^{+ \infty}
I_l (s) e^{ \frac{ \beta l \bar{\omega}_0}{2}}
[ 1 - n_F(\omega - l \bar{\omega}_0)] g(\omega - l \bar{\omega}_0) +
\nonumber \\
&& 2 \pi e^{-S_T}  \sum_{l(\neq 0)- \infty}^{+ \infty} I_l (s) 
e^{ \frac{ \beta l \bar{\omega}_0}{2}}
n_F(\omega + l \bar{\omega}_0) g(\omega+ l \bar{\omega}_0) +
\nonumber \\
&& 2 \pi e^{-S_T} [I_0(s) -1 ] g(\omega).
\label{21r}
\end{eqnarray}
The first term represents the purely polaronic band contribution and shows a 
delta behavior. The remaining terms provide the incoherent contribution 
and spread over a wide energy range. 
Thus we get the renormalized density of states 
\begin{eqnarray}
 N(\omega)=&&
\frac{1}{2 \pi} \left( \frac{1}{N} \right)
\sum_{{\bf k}} A( {\bf k}, \omega),
\label{22r}
\end{eqnarray}
where the spectral function $A( {\bf k}, \omega)$ is given by eq.(\ref{21r}).
We note that the sum rule 
$ \int_{- \infty}^{+ \infty} \frac{d \omega}{2 \pi} A( {\bf k}, \omega) =1 $
is satisfied and $ N( \omega)$ is normalized to unity.

So we can evaluate the renormalized electron distribution function

\begin{eqnarray}
n^{ren}_{{\bf k}}=&&
\int_{- \infty}^{+ \infty} \frac{d \omega}{2 \pi} 
A( {\bf k}, \omega) n_F ( \omega )=  
e^{-S_T} n_F( \xi_{{\bf k}} )+
\nonumber \\
&& e^{-S_T} \sum_{l(\neq 0)- \infty}^{+ \infty} I_l (s) 
e^{ \frac{ \beta l \bar{\omega}_0}{2}}
\int_{-\infty}^{\infty} d\xi g(\xi) [1-n_{F}(\xi)]
n_F ( \xi +l \bar{\omega}_0 ) + 
\nonumber \\
&& e^{-S_T} \sum_{l(\neq 0)- \infty}^{+ \infty} I_l (s) 
e^{ \frac{ \beta l \bar{\omega}_0}{2}}
\int_{-\infty}^{\infty} d\xi g(\xi) n_{F}(\xi)
n_F ( \xi -l \bar{\omega}_0 ) + 
\nonumber \\
&& e^{-S_T} [ I_0 (s) -1 ] 
\int_{-\infty}^{\infty} d\xi g(\xi) n_{F}(\xi).
\label{24r}
\end{eqnarray}
At zero temperature it is found 
$ n^{ren}( \mu -\delta )-n^{ren}( \mu +\delta )= e^{-S_T} $, so the factor 
$e^{-S_T}$ determines the jump in the Fermi distribution function. \cite{alex}
For low temperatures and $x>0.25$, $e^{-S_T}$ is about 0.9, 
therefore the spectral properties indicate a polaronic Fermi-liquid state in 
the metallic ferromagnetic phase. 

On the other hand, for the small polaron excitations and high temperatures, 
we expect that the spectral properties are dominated only by the incoherent 
term.
It is well known \cite{mahan} that in this limit the 
spectral function (\ref{21r}) consists of a sum of delta functions on the 
points 
$n \omega_0$. This is caused by the choice of a dispersionless phonon 
frequency. 
To overcome this drawback, there are two ways of proceeding: the first is to 
introduce a phononic dispersion relation \cite{12}, whereas the second 
is to make a high temperature expansion.\cite{loosh}
In this paper we concentrate on the latter approach performing in 
eq.(\ref{17r}) the following approximation \cite{loosh}  

\begin{equation}
e^{-S_T} 
\exp
\left\{s\cosh\left[\bar{\omega}_0\left(\tau-\frac{\beta}{2}\right)\right]
\right\} 
\simeq
\exp(- \bar{s} \tau)
\exp \left( \frac{\bar{s}}{\beta} \tau^2 \right) 
\label{25r}
\end{equation}
where $\bar{s}$ is

\begin{equation}
\bar{s}= \frac{2 f^2 g^2 e^{-4 \alpha} \bar{\omega}_0^2 \beta }   
{4 \sinh \left( \frac{ \beta \bar{\omega}_0 }{2} \right) } 
\simeq
g^2 \omega_0. 
\label{26r}
\end{equation}
This expansion is valid as long as
$ \left( \frac{ \beta \bar{s}}{12} \right) 
 \left( \frac{ \beta \bar{\omega}_0}{4} \right)^2 <1 $, 
hence $T>0.3 \omega_0 $.
In this limit the Green's function becomes

\begin{eqnarray}
{ \mathcal G} \left( {\bf k},i \omega_n \right)=&&
e^{-S_T} { \mathcal G}^{(0)} \left( { \bf k}, i \omega_n \right)  
- e^{-S_T} \left( \frac{1}{N} \right) \sum_{\bf{k}_1}
{ \mathcal G}^{(0)} \left( {\bf k}_1,i \omega_n \right)+
\nonumber \\
&& \left( - \frac{i}{2} \right) 
\sqrt{ \frac{ \pi \beta}{ \bar{s}}}
\left( \frac{1}{N} \right) \sum_{\bf{k}_1}
\left( 1-n_{{\bf k}_1} \right) 
W \left[ \frac{1}{2} \sqrt{ \frac{ \beta}{ \bar{s}} }
\left( i \omega_n -\xi_{{\bf k}_1} - \bar{s}  \right) \right]+
\nonumber \\
&& \left(- \frac{i}{2} \right) 
\sqrt{ \frac{\pi \beta}{\bar{s}}}
\left( \frac{1}{N} \right) \sum_{{\bf k}_1}
n_{ {\bf k}_1} W \left[ \frac{1}{2} 
\sqrt{ \frac{ \beta}{ \bar{s}} }
\left( i \omega_n -\xi_{{\bf k}_1} + \bar{s}  \right) \right] 
\label{27r}
\end{eqnarray}
where $W(z)$ is

\begin{equation}
W(z)=e^{-z^2}
\left[
1 + \frac{2i}{\sqrt{\pi}} \int_0^z d\zeta e^{\zeta^2}
\right].
\label{28r}
\end{equation}

In the small polaron case, we get $ \xi_{{\bf k}} \simeq -g^2 \omega_0 -\mu $. 
The incoherent term yields 

\begin{eqnarray}
A \left( {\bf k}, \omega \right)
\simeq && A(\omega)=
\nonumber \\
&& \left[ 1-n_F \left( -g^2 \omega_0 - \mu \right) \right] 
\sqrt{ \frac{ \pi \beta}{ \bar{s}}}
\exp \left[  - \frac{\beta}{4 \bar{s}} 
( \omega +g^2 \omega_0 + \mu - \bar{s} )^2  \right] +
\nonumber \\
&& n_F (-g^2 \omega_0 - \mu ) 
\sqrt{ \frac{ \pi \beta}{ \bar{s}}}
\exp \left[  - \frac{\beta}{4 \bar{s}} 
(\omega +g^2 \omega_0 + \mu + \bar{s} )^2  \right].
\label{29r}
\end{eqnarray}
We stress that this spectral function represents the envelope of the delta 
functions obtained by eq.(\ref{21r}) for small polaron excitations in the 
limit of high temperatures.
Furthermore, using the spectral function (\ref{29r}), we can derive the 
renormalized density of states 

\begin{equation}
N(\omega)= \frac {A (\omega)} {2 \pi}
\label{29ar}
\end{equation}
that turns out normalized to unity.

In Fig.2 we report the spectral function for two different temperatures at 
$x=0.3$. 
At $T=0.05 \omega_0$, the spectral function is deduced by eq.(\ref{21r}) and 
the case $\xi_{\bf{k}}=0$ is considered. It is apparent that in the 
large polaron ferromagnetic phase the weight of the coherent term is prevalent.
For  $T=0.47 \omega_0$, in the paramagnetic phase, the spectral
function is provided by eq.(\ref{29r}) and it is determined by
the incoherent dynamics of the small polaron excitations. It is made up of
two bands peaked approximatively around $-g^2 \omega_0$ and $g^2 \omega_0$, 
whose different heights are due to the fixed doping. At high temperatures the 
spectral function is characterized by a large depression in correspondence of 
$\mu$.  
With decreasing temperature and fixed doping, a transfer of spectral
weight occurs crossing the regime of coexisting phases. 
To stress this important point, in Fig.3 we present the renormalized density 
of states for $x=0.3$ at different temperatures. In the large polaron 
ferromagnetic phase, $N(\omega)$ is given by eq.(\ref{22r}), in the small 
polaron paramagnetic phase by eq.(\ref{29r}), in the regime of coexisting 
phases by the two preceding densities by means of the eq.(\ref{14r}). We 
stress that $T_c$ denotes the ferromagnetic transition temperature which 
coincides with the beginning of formation of large polaron domains. 
For $x=0.3$, $T_c$ is about $0.46 \omega_0$.
For low temperatures incoherent processes of phonon creation and annihilation 
lead to a depression in the density of states around the chemical potential 
$\mu$.
When the temperature increases, the effect of band renormalization driven 
by $e^{-S_T}$ is stronger than this decrease of electronic states at $\mu$. 
Since the density of states is normalized, the net effect is a rise of states 
at $\mu$ (compare graphs at $T=0.28 T_c$ and $T=0.65 T_c$).
However, when the phase separation regime takes place 
(starting from $T=0.65T_c$), 
the formation of paramagnetic insulating domains in the metallic background 
causes a robust pseudogap in the excitation spectrum. The subtraction of 
levels below $\mu$ is strongly temperature dependent and, 
as we see in the inset of Fig.3, it finishes at $T_c$. 
Therefore the pseudogap features show a direct relationship 
with the metal-insulator transition and the coherent-incoherent crossover of 
charge dynamics.         
Above $T_c$ the density of states at $\mu$ is low and with increasing 
temperature it is nearly constant. 
It is important to note that these results are compatible with 
photoemission experiments.
Indeed the density of states at $\mu$ in the inset of Fig.3 bears a strong 
resemblance with the spectral weights reported in refs \cite{dessau1,dessau2}. 

Finally it is interesting to study the pseudogap formation fixing the 
temperature and varying the hole concentration $x$ (see Fig.4). 
At $T=0.39 \omega_0$, the system is in the insulating paramagnetic phase from 
$x=0$ to $x=0.17$, in the regime of the insulating paramagnetic and metallic 
ferromagnetic coexisting phases from $x=0.17$ to $x=0.33$, in the metallic 
ferromagnetic phase from $x=0.33$. Increasing $x$, holes are added to the 
system and ferromagnetic large polaron domains are created yielding pseudogap 
features.
As soon as the phase separation regime is set up at $x=0.17$ (inset of Fig.4), 
electronic levels are occupied in the energy range around the chemical 
potential $\mu$. Thus, the charge carriers prefer to be located in the 
metallic ferromagnetic domains in order to increase the kinetic energy gain.

\section{Damping}

In this section we deal with the polaron self-energy. This quantity allows 
to determine the scattering rate of particles affected by lattice distortions 
and spin fluctuations. The latter, in turn, will play an essential role in 
the infrared absorption calculations.

Retaining only the dominant autocorrelation terms at the second step of 
iteration \cite{schna,loos1,loos2,kada,fehske}, we can derive the self-energy

\begin{eqnarray}
\Sigma^{(2)} \left( {\bf k}, i \omega_n \right)=  
\varepsilon_{\bf k}+
\left( \frac{1}{N} \right) \sum_{\bf{k}_1}
\left( \frac{1}{\beta} \right) \sum_{m}
{\mathcal G}^{0}  \left( {\bf k}_1,i \omega_m \right)
\int_0^{\beta} d\tau e^{(i \omega_n -i \omega_m ) \tau}
[f_1(\tau)+f_2(\tau)]
\label{34r}
\end{eqnarray}
where $\varepsilon_{\bf k}$ is given by eq.(\ref{11ur}), $f_1(\tau)$ is

\begin{eqnarray}
f_1(\tau)=Z t^2 e^{ -2 S_T} 
\left\{ 
\left\langle \left( \frac{S_0+1/2}{2 S+1} \right)^2  \right\rangle
e^{ z\cosh \left[\bar{\omega}_0\left(\tau-\frac{\beta}{2}\right)\right] }
- \left\langle \left( \frac{S_0+1/2}{2 S+1} \right)  \right\rangle^2 
\right\}
\label{35r}
\end{eqnarray}
with

\begin{equation}
z=4 g^2 f^2 e^{ -4 \alpha } [N_0(N_0+1)]^{ \frac{1}{2}}   
\label{36r}
\end{equation}
and $f_2(\tau)$ is 

\begin{eqnarray}
f_2(\tau)=2 g^2 \omega_0^2 e^{ 4 \alpha } (1-f)^2 
[N_0(N_0+1)]^{ \frac{1}{2}} 
\cosh 
\left[ 
\bar{\omega}_0 \left( \tau- \frac{\beta}{2} \right) 
\right]. 
\label{37r}
\end{eqnarray}
Making the analytic continuation $i \omega_n \rightarrow \omega +i\delta$, we 
can calculate the scattering rate 
	
\begin{equation}
\Gamma( {\bf k})=\tilde{\Gamma} ( {\bf k},\omega = \xi_{{\bf k}}  )
=-2 \Im \Sigma^{(2)}_{ret} 
\left( {\bf k}, \omega= \xi_{{\bf k}}  \right).
\label{42r}
\end{equation}
We realize that the dependence of $\Gamma( {\bf k})$ on $ {\bf k }$ is 
due only to $ \xi _{{\bf k}} $. Hence, performing the expansion 
(\ref{20r}) into the series of multiphonon processes, we can express the 
scattering rate in the following way 

\begin{equation}
\Gamma( {\bf k})=\Gamma(\xi_{{\bf k}})= 
\Gamma_{1-phon}(\xi_{{\bf k}})+ \Gamma_{multi-phon}(\xi_{{\bf k}}) + 
\Gamma_{Spin-Fluct}(\xi_{{\bf k}})
\label{44r}
\end{equation}
where $\Gamma_{1-phon}$ is the contribution due to single phonon
processes only 

\begin{eqnarray}
\Gamma_{1-phon}(\xi_{{\bf k}})=&&
2 Z t^2 e^{ -2 S_T} 
\left\langle \left( \frac{S_0+1/2}{2 S+1} \right)^2  \right\rangle
I_1(z) 
\sinh \left( \frac{\beta \bar{\omega}_0 }{2} \right)
g_{1,l=1}(\xi_{{\bf k}})+
\nonumber \\
&& g^2 \omega_0^2 e^{ 4 \alpha } (1-f)^2 g_2(\xi_{{\bf k}}),
\label{45r}
\end{eqnarray}
$\Gamma_{multi-phon}$ represents the scattering rate by 
multiphonon processes 

\begin{equation}
\Gamma_{multi-phon}(\xi_{{\bf k}})=
2 Z t^2 e^{ -2 S_T} 
\left\langle \left( \frac{S_0+1/2}{2 S+1} \right)^2  \right\rangle
\sum_{l=2}^{+ \infty} I_l(z) 
\sinh \left( \frac{\beta \bar{\omega}_0 l}{2} \right)
g_{1,l}(\xi_{{\bf k}})
\label{46r}
\end{equation}
and $\Gamma_{Spin-Fluct}$ denotes the damping term by spin fluctuations
  
\begin{equation}
\Gamma_{Spin-Fluct}(\xi_{{\bf k}})=
Z t^2 e^{ -2 S_T} 
\left[ 
I_0(z)
\left\langle \left( \frac{S_0+1/2}{2 S+1} \right)^2  \right\rangle
- \left\langle \left( \frac{S_0+1/2}{2 S+1} \right)  \right\rangle^2 
\right]
B(\xi_{{\bf k}}).
\label{47r}
\end{equation}
Here $g_{1,l} $ reads 

\begin{equation}
g_{1,l}(\xi_{{\bf k}})=
\left[
N_0 (l \bar{\omega}_0 )+n_F(\xi_{{\bf k}} +l \bar{\omega}_0)
\right]
B(\xi_{{\bf k}} +l \bar{\omega}_0 )+
\left[
N_0 (l \bar{\omega}_0 )+1-n_F(\xi_{{\bf k}} -l \bar{\omega}_0)  
\right]
B(\xi_{{\bf k}} -l \bar{\omega}_0 )
\end{equation}
and $g_2 $
 
\begin{equation}
g_2(\xi_{{\bf k}})=
\left[
N_0 +n_F(\xi_{{\bf k}} +\bar{\omega}_0)
\right]
B(\xi_{{\bf k}} +\bar{\omega}_0 )+
\left[
N_0 +1-n_F(\xi_{{\bf k}} -\bar{\omega}_0)  
\right]
B(\xi_{{\bf k}} - \bar{\omega}_0 ).
\label{49r}
\end{equation}
The function $B(x)$ is defined as
\begin{equation}
B(x)=2 \pi g(x) 
\end{equation}
where $g(x)$ is the density of states (\ref{12r}).

We stress that the decomposition of the scattering rate in three distinct terms
has been introduced in order to simplify the analysis of our results.
Furthermore we observe that the scattering rate for small 
polaron excitations can be correctly calculated only within a self-consistent 
treatment. \cite{loos1,loos2} 
This can be carried out substituting in eq.(\ref{34r}) ${\mathcal G}^{(0)}$ for
a new Green's function $\tilde{ {\mathcal G} }$.
Using the Lehmann representation we can write

\begin{equation}
\tilde{ {\mathcal G} } ( {\bf k},i \omega_n )=
\int_{- \infty}^{+ \infty} \frac{ d \omega } {2 \pi} 
\frac
{ \tilde{ A }( {\bf k}, \omega )}
{i \omega_n - \omega }  
\label{38r}
\end{equation} 
where the spectral function $\tilde{ A }$ is assumed to be

\begin{equation}
\tilde{ A }( {\bf k}, \omega )=
\frac
{\Gamma( {\bf k})}
{ [\Gamma( {\bf k})]^2/4 +(\omega - \xi_{{\bf k}} )^2   }.
\label{43r}
\end{equation}
So, employing this self-consistent procedure, we get the scattering rate 
$\Gamma^{(sc)}(\xi_{{\bf k}})$

\begin{equation}
\Gamma^{(sc)}(\xi_{{\bf k}})= \Gamma^{(sc)}_{1-phon}(\xi_{{\bf k}})+ 
\Gamma^{(sc)}_{multi-phon}(\xi_{{\bf k}}) + 
\Gamma^{(sc)}_{Spin-Fluct}(\xi_{{\bf k}})
\label{51r}
\end{equation}
where $\Gamma^{(sc)}_{1-phon}$, 
$\Gamma^{(sc)}_{multi-phon}$
and $\Gamma^{(sc)}_{Spin-Fluct}$ 
are calculated by eqs (\ref{45r}),(\ref{46r}),(\ref{47r}), respectively, 
substituting the function $B(x)$ for $B^{(sc)}(x)$

\begin{equation}
B^{(sc)}(x)= \left( \frac{1}{N} \right) \sum_{{\bf k}_1} 
\tilde{ A }( {\bf k}_1,x )=
\int_{- \infty}^{+ \infty} d \xi_1 g(\xi_1) 
\tilde{ A }( \xi_1,x ).
\label{48r}
\end{equation}
Since the spectral function $\tilde{ A }$ (\ref{43r}) depends  on 
$\Gamma$, the eq.(\ref{51r}) allows to evaluate self-consistently the damping 
due to polaron formation and spin fluctuations.
Clearly when in the right-hand side of eq.(\ref{51r}) the limit 
$\Gamma \rightarrow 0$ is made, the equation (\ref{44r}) is obtained.

In Fig.5a we report the scattering rate $\Gamma$ of large polaron excitations 
for different temperatures. Because of the augmented number of phonons and the
enhanced role of spin fluctuations with rising temperature, the quantity 
$\Gamma$ increases. 
At low temperatures we notice that the scattering rate is zero for energies 
within $\omega_0$ of the chemical potential $\mu$. This means that 
at these temperatures the spin fluctuations are too feeble to scatter 
electrons and the main mechanism of energy loss is the emission or the 
absorption of phonons with frequency $\omega_0$ (one phonon processes 
are prevalent). Thus the behavior of the scattering rate is explained by the 
fact that the quasi-particle excitations within $\omega_0$ of the Fermi energy
cannot lose energy  because of the Fermi statistics.\cite{mahan}

In Fig.5b we concentrate on the components of the scattering rate at a fixed
temperature ($T=0.65 T_c$). It is confirmed that also at higher temperatures 
the single phonon emission and absorption give an important contribution
to the quantity $\Gamma$. Nevertheless we stress that the spin fluctuations 
are effective to enhance the scattering rate just in the important energy range
around the chemical potential $\mu$.  
Finally the multi-phonon contribution is dominated by two-phonon processes 
and it is negligible.

We stress that for the small polaron excitations the quantity $\Gamma$ 
decreases when $T$ increases resembling the behavior of $t_{eff}$. 
We note, however, that, for $T>0.3 \omega_0$, $\Gamma$ is always larger 
than $t_{eff}$, making clear that the electronic states lose their individual 
characteristics and the electron motion is predominantly a diffusive process.

The introduction of the damping allows to improve the 
approximations of calculation for the spectral properties: indeed we 
can determine the spectral function  

\begin{eqnarray}
A( {\bf k}, \omega) = && e^{-S_T} \tilde{ A }( {\bf k}, \omega )+
\nonumber \\
&& 2  e^{-S_T} \sum_{ l=1 }^{ + \infty}
I_l (s) \sinh \left(  \frac{ \beta l \bar{\omega}_0}{2} \right)
\left( \frac{1}{N} \right) \sum_{\bf{k}_1}
\left[
N_0 (l \bar{\omega}_0 )+n_F(\omega +l \bar{\omega}_0)
\right]
 \tilde{ A }( {\bf k}_1, \omega +l \bar{\omega}_0 )+
\nonumber \\
2 &&  e^{-S_T} \sum_{ l=1 }^{ + \infty}
I_l (s) \sinh \left(  \frac{ \beta l \bar{\omega}_0}{2} \right)
\left( \frac{1}{N} \right) \sum_{\bf{k}_1}
\left[
N_0 (l \bar{\omega}_0 )+1-n_F(\omega -l \bar{\omega}_0)  
\right]
\tilde{ A }( {\bf k}_1, \omega +l \bar{\omega}_0 )+
\nonumber \\
&& e^{-S_T} [I_0(s) -1 ] \left( \frac{1}{N} \right) \sum_{\bf{k}_1}
 \tilde{ A }( {\bf k_1}, \omega ).
\label{52r}
\end{eqnarray}
In the limit $\Gamma \rightarrow 0$, the result of eq.(\ref{21r}) is recovered.
We have checked that the quantity $\Gamma$ does not change the spectral 
properties in a considerable manner, although it allows to eliminate the 
delta behavior in the expression (\ref{21r}).

On the other hand, as mentioned, the damping is essential for determining the 
optical properties of the system.

\section{Optical properties}

In this section we focus our attention on the optical properties.

To determine the linear response to an external field of frequency $\omega$,
we derive the conductivity tensor $\sigma_{\alpha,\beta}$ by means of the Kubo
formula 
\begin{equation}
\sigma_{\alpha,\beta}(\omega)=\left( \frac {i e^2 }{\omega} \right)
 <\hat{T}_{\alpha,\beta}>+ 
\left( \frac{i}{\omega} \right) \Pi_{\alpha,\beta}^{ret}(\omega)
\label{53r}
\end{equation}
where $\hat{T}_{\alpha,\beta}$ is the kinetic energy and 
$\Pi_{\alpha,\beta}^{ret}(\omega)$ is the retarted current-current correlation
function.
The transport of electrons will be supposed to take place between the 
equivalent $nn$ sites of the cubic lattice, hence the tensor 
$\sigma_{\alpha,\beta}$ is assumed to be diagonal with mutually equal elements 
$\sigma_{\alpha,\alpha}$.
To calculate the infrared absorption, we need only the real part of the 
conductivity

\begin{equation}
\Re \sigma_{\alpha,\alpha}(\omega)= 
- \frac{ \Im \Pi_{\alpha,\alpha}^{ret}(\omega) }{\omega}.
\label{54r}
\end{equation}
Therefore our problem reduces to evaluate the current-current correlation 
function that in Matsubara frequencies is defined as    
\begin{equation}
\Pi_{\alpha,\alpha}(i \omega_n)=  \left(- \frac{1}{N} \right)
\int_0 ^{\beta} e^{i \omega_n \tau} 
\langle T_{\tau} j^{\dagger}_{\alpha}(\tau) j_{\alpha}(0) \rangle 
\label{55r}
\end{equation}
where the current operator $j_{\alpha}$ suitable for the Hamiltonian is
(\ref{1r})

\begin{equation}
j_{\alpha}=ite \sum_{i,\delta}  
\left( \frac{S_0^{i+\delta \hat{\alpha},i}+1/2}{2 S+1} \right)
c^{\dagger}_{i+\delta \hat{\alpha}} c_i
\label{56r}
\end{equation}
and 
$ j^{\dagger}_{\alpha}(\tau)=e^{H \tau} j^{\dagger}_{\alpha} e^{-H \tau} $ .  
Performing the two canonical transformations (\ref{2r},\ref{3r}) and 
making the decoupling of the correlation function (\ref{55r}) in the electron,
phonon and spin terms through the introduction of $H_{test}$ (\ref{9r}), we get

\begin{equation}
\Pi_{\alpha,\alpha}(i \omega_n)=e^2 t^2 
\left(- \frac{1}{N} \right) \sum_{i,\delta} \sum_{i^{\prime},\delta^{\prime}}
( \delta \cdot \delta^{\prime} )
\int_0 ^{\beta} e^{i \omega_n \tau} 
\Delta (i,i^{\prime},\delta,\delta^{\prime})
\langle T_{\tau} \bar{c}_i^{\dagger}(\tau) \bar{c}_{i+\delta \hat{\alpha}}
(\tau)
c_{i^{\prime}+\delta^{\prime} \hat{\alpha}}^{\dagger} c_{i^{\prime}} \rangle_t
\label{57r}
\end{equation}
where the function $\Delta$

\begin{equation}
\Delta (i,i^{\prime},\delta,\delta^{\prime}) = 
\Phi\left( i,i^{\prime},\delta,\delta^{\prime} \right) 
\Upsilon (i,i^{\prime},\delta,\delta^{\prime}) 
\label{58r}
\end{equation}
denotes the product of the phonon correlation function $\Phi$ 

\begin{equation}
\Phi\left( i,i^{\prime},\delta,\delta^{\prime} \right) =
\langle T_{\tau} \bar{X}_i^{\dagger}(\tau) \bar{X}_{i+\delta \hat{\alpha}}
(\tau)
X_{i^{\prime}+\delta^{\prime} \hat{\alpha}}^{\dagger} X_{i^{\prime}} \rangle_t
\label{59r}
\end{equation}
times the spin correlation function $\Upsilon$

\begin{equation}
\Upsilon \left( i,i^{\prime},\delta,\delta^{\prime} \right) =
\left\langle
\left( \frac{S_0^{i,i+\delta \hat{\alpha}}+1/2}{2 S+1} \right)
\left( \frac{S_0^{i^{\prime}+\delta^{\prime} \hat{\alpha},i^{\prime}}+1/2}
{2 S+1} \right)
\right\rangle_t.
\label{61r}
\end{equation}
We calculate the electron correlation function

\begin{eqnarray}
&& \langle T_{\tau} \bar{c}_i^{\dagger}(\tau) \bar{c}_{i+\delta \hat{\alpha}}
(\tau)
c_{i^{\prime}+\delta^{\prime} \hat{\alpha}}^{\dagger} c_{i^{\prime}} \rangle_t=
\nonumber \\
\left(- \frac{1}{N^2} \right) \sum_{{\bf k},{\bf k}_1}
e^{ i {\bf k} \cdot {\bf R}_i  }
&& e^{ -i {\bf k}_1 \cdot ( {\bf R}_i + \delta \hat{\alpha} )  } 
e^{ i {\bf k}_1 \cdot ( {\bf R}_{i^{\prime}}  + \delta^{\prime}   
\hat{\alpha} )  }e^{ -i {\bf k} \cdot {\bf R}_{i^{\prime}}  }
{ \mathcal G}^{(0)} ( {\bf k},-\tau )
{ \mathcal G}^{(0)} ( {\bf k}_1,\tau )
\label{62r}
\end{eqnarray}
and we separate $\Delta$ into two terms

\begin{eqnarray}
\Delta (i,i^{\prime},\delta,\delta^{\prime}) = && 
\Phi\left( i,i^{\prime},\delta,\delta^{\prime} \right) 
\Upsilon (i,i^{\prime},\delta,\delta^{\prime})=
\nonumber \\
&&
\left[ \langle X_i^{\dagger} X_{i+\delta \hat{\alpha}}\rangle_t \right]^2
\left\langle
\left( \frac{S_0^{i,i+\delta \hat{\alpha}}+1/2}{2 S+1} \right)
\right\rangle^2_t+
\nonumber \\
&& \left\{
\Phi\left( i,i^{\prime},\delta,\delta^{\prime} \right) 
\Upsilon (i,i^{\prime},\delta,\delta^{\prime})-
\left[ \langle X_i^{\dagger} X_{i+\delta \hat{\alpha}}\rangle_t \right]^2
\left\langle
\left( \frac{S_0^{i,i+\delta \hat{\alpha}}+1/2}{2 S+1} \right)
\right\rangle^2_t
\right\}
\label{63r}
\end{eqnarray}
hence
\begin{eqnarray}
\Delta (i,i^{\prime},\delta, && \delta^{\prime}) =  
\nonumber \\
&& e^{ -2 S_T} 
\left\langle \left( \frac{S_0+1/2}{2 S+1} \right) \right\rangle^2+
\left[
\Phi\left( i,i^{\prime},\delta,\delta^{\prime} \right) 
\Upsilon (i,i^{\prime},\delta,\delta^{\prime})-
e^{ -2 S_T} 
\left\langle \left( \frac{S_0+1/2}{2 S+1} \right) \right\rangle^2
\right].
\label{64r}
\end{eqnarray}

To derive the optical properties, the role of the damping $\Gamma$ of the 
particle motion is fundamental. Furthermore the defect scattering rate 
$\Gamma_0$ is added to the quantity $\Gamma$ calculated in the previous 
section. This scattering rate $\Gamma_0$ is assumed to be a small 
perturbation, thus it has been fixed equal to $0.1 \omega_0$.
The effect of the damping can enter our calculation substituting in 
eq.(\ref{62r}) $ {\cal G}^{(0)}$ for $ \tilde{{\cal G}}$ given by 
eq.(\ref{38r}).  

Considering the two terms of eq.(\ref{64r}), the correlation function can be 
written as

\begin{equation}
\Pi_{\alpha,\alpha}(i \omega_n)=
\Pi^{(1)}_{\alpha,\alpha}(i \omega_n)+
\Pi^{(2)}_{\alpha,\alpha}(i \omega_n).
\label{65r}
\end{equation}
The first term reads

\begin{equation}
\Pi^{(1)}_{\alpha,\alpha}(i \omega_n)=
4 e^2 t^2 e^{ -2 S_T} 
\left\langle \left( \frac{S_0+1/2}{2 S+1} \right) \right\rangle^2
\left( \frac{1}{N} \right) \sum_{{\bf k}} \sin^2(k_{\alpha})
\int_0 ^{\beta} e^{i \omega_n \tau}
 \tilde{{\cal G}} ({\bf k},-\tau)   
 \tilde{{\cal G}} ({\bf k},\tau)   
\label{66r}
\end{equation}
and the second one is obtained retaining only the main autocorrelation 
term $i=i^{\prime}$ and $ \delta=\delta^{\prime}$ 

\begin{equation}
\Pi^{(2)}_{\alpha,\alpha}(i \omega_n)=
\left( \frac {2 e^2 } {Z} \right)  
\left\langle \left( \frac{S_0+1/2}{2 S+1} \right) \right\rangle^2
\left( \frac{1}{N^2} \right) \sum_{{\bf k},{\bf k_1}} 
\int_0 ^{\beta} e^{i \omega_n \tau}
\tilde{{\cal G}} ({\bf k},-\tau)
\tilde{{\cal G}} ({\bf k}_1,\tau)
f_1(\tau)       
\label{67r}
\end{equation}
where $f_1(\tau)$ is given by eq.(\ref{35r}).
We stress that substituting $ {\cal G}^{(0)}$ for $ \tilde{{\cal G}}$  
is necessary to have a non vanishing $ \Pi^{(1)}_{\alpha,\alpha}
(i \omega_n)   $.
We perform the analytic continuation $i \omega_n \rightarrow \omega +i\delta$, 
and, clearly, the conductivity can be expressed as sum of two terms 

\begin{equation}
\Re \sigma_{\alpha,\alpha}(\omega)= 
- \frac{ \Im \left[ \Pi_{\alpha,\alpha}^{ret (1)}(\omega)+ 
\Pi_{\alpha,\alpha}^{ret (2)}(\omega) \right]} {\omega}
=\Re \sigma^{(band)}_{\alpha,\alpha}(\omega)
+ \Re \sigma^{(incoh)}_{\alpha,\alpha}(\omega).
\label{68r}
\end{equation}
The first term $\Re \sigma^{(band)}_{\alpha,\alpha}$ represents the 
band conductivity. Here the charge transfer is affected by the damping due 
to interactions with the lattice and spin fluctuations, but it is not 
accompanied by processes changing the number of phonons. 
On the other hand, the second term $ \Re \sigma^{(incoh)}_{\alpha,\alpha}$ 
in eq. (\ref{68r}) derives from inelastic scattering processes of emission 
and absorption of phonons and it is marked by the apex ``incoherent'' since 
the particles lose their phase coherence because of this phonon-assisted 
hopping. As in the spectral properties, we stress the appearance of two 
physically distinct contributions. 

In appendix we report the details of calculation for the optical conductivity.
The band conductivity is derived as 

\begin{equation}
\Re \sigma^{(band)}_{\alpha,\alpha}(\omega)= 
\left( \frac{ 4 e^2 t^2}{\omega} \right) e^{ -2 S_T} 
\left\langle \left( \frac{S_0+1/2}{2 S+1} \right) \right\rangle^2
\int_{- \infty}^{+ \infty} d \xi 
 [n_F(\xi-\omega)-n_F(\xi)]
\tilde{C}(\xi,\omega) h(\xi)
\label{130ar}
\end{equation} 
where $\tilde{C}(\xi,\omega)$ is    

\begin{equation}
\tilde{C}(\xi,\omega)=  
\frac{ \Gamma(\xi) }
{ \Gamma^2(\xi)+\omega^2 }
\label{131ar}
\end{equation} 
and $h(\xi)$ reads

\begin{equation}
h(\xi)=\left( \frac{1}{N} \right) \sum_{{\bf k}} \sin^2(k_{\alpha})
\delta( \xi - \xi_{{\bf k}} ).
\label{132ar}
\end{equation} 
The latter term of the conductivity becomes

\begin{eqnarray}
\Re \sigma^{(incoh)}_{\alpha,\alpha}(\omega)= && 
\left( \frac{ 2 e^2 t^2}{\omega} \right) e^{ -2 S_T} 
\left\langle \left( \frac{S_0+1/2}{2 S+1} \right)^2 \right\rangle
\int_{- \infty} ^{+ \infty} d \xi 
\int_{- \infty} ^{+ \infty} d \xi_1  
g(\xi) g(\xi_1) R(\xi,\xi_1,\omega)+
\nonumber \\
&& \left( \frac{ 2 e^2 t^2}{\omega} \right) e^{ -2 S_T} 
\left[ I_0(z) \left\langle \left( \frac{S_0+1/2}{2 S+1} \right)^2  
\right\rangle
- \left\langle \left( \frac{S_0+1/2}{2 S+1} \right)  \right\rangle^2  \right]
\times
\nonumber \\
&& \times
\int_{- \infty} ^{+ \infty} d \xi 
\int_{- \infty} ^{+ \infty} d \xi_1  
g(\xi) g(\xi_1)  [n_F(\xi-\omega)-n_F(\xi)]
C(\xi,\xi_1,\omega) 
\label{134ar}
\end{eqnarray}
where $g(\xi)$ is the density of states (\ref{12r}) and the function
$ R(\xi,\xi_1,\omega) $ is given by 

\begin{equation}
 R(\xi,\xi_1,\omega)=
2\sum_{l=1}^{+ \infty} I_l(z) 
\sinh \left( \frac{\beta \bar{\omega}_0 l}{2} \right)
\left[ 
J_l(\xi,\xi_1,\omega)+
H_l(\xi,\xi_1,\omega)
\right].
\label{135ar}
\end{equation}
We notice that $J_l( \xi,\xi_1,\omega )$ 

\begin{eqnarray}
J_l( \xi,\xi_1,\omega )=
C( \xi,\xi_1,\omega+l\bar{\omega}_0 ) 
[n_F(\xi-l\bar{\omega}_0-\omega)-
n_F(\xi-l\bar{\omega}_0)]
\left[ N_0(l \bar{\omega}_0)+n_F(\xi) \right]
\label{136ar}
\end{eqnarray}
and $H_l( \xi,\xi_1,\omega )$ 
 
\begin{eqnarray}
H_l( \xi,\xi_1,\omega )=
C( \xi,\xi_1,\omega-l\bar{\omega}_0 )
[n_F(\xi+l\bar{\omega}_0-\omega)-
n_F(\xi+l\bar{\omega}_0)]
\left[ N_0(l \bar{\omega}_0)+1-n_F(\xi) \right]
\label{137ar}
\end{eqnarray} 
describe phonon absorption and emission processes, respectively. 

In the limit of high temperatures ($T>0.39 \omega_0$) and for small polaron 
excitations, the incoherent absorption is prevalent. In this case the 
conductivity consists of a sum of narrow Lorentzian functions
centred on the points $n \omega_0$ respectively. \cite{mahan,loosh} 
We perform an expansion suitable for high temperatures and in appendix we 
derive the approximate expression of the conductivity 

\begin{eqnarray}
&& \Re \sigma_{\alpha,\alpha}(\omega)= 
\nonumber \\
&&
\left( \frac{e^2 t^2}{\omega} \right)  
\left\langle \left( \frac{S_0+1/2}{2 S+1} \right)^2 \right\rangle
\sqrt{ \frac{ \pi \beta}{ \bar{z}}}
\left\{
\exp \left[  - \frac{\beta}{4 \bar{z}} 
( \omega - \bar{z} )^2  \right] -
\exp \left[  - \frac{\beta}{4 \bar{z}} 
(\omega + \bar{z} )^2  \right]
\right\}
\rho ( 1- \rho )
\label{60ar}
\end{eqnarray}
where $\rho$ is the electron concentration.

We have checked the internal consistency of our approach by means of the
sum rule
\begin{equation}
\int_0^{\infty} d\omega \Re \sigma_{\alpha,\alpha}(\omega)=
- \frac{\pi}{2} e^2 <\hat{T}_{\alpha,\alpha}>   
\label{560r}
\end{equation}
where $\hat{T}_{\alpha,\alpha}$ is the component of the kinetic energy.
The left-hand side of eq.(\ref{560r}) is calculated by using the conductivity 
spectra, whereas the mean value $<\hat{T}_{\alpha,\alpha}>$ can be evaluated 
by performing the two canonical transformations (\ref{2r},\ref{3r}) and making 
the decoupling in the electron, phonon and spin terms through the introduction
of $H_{test}$ (\ref{9r}). We get

\begin{equation}
<\hat{T}_{\alpha,\alpha}>= \frac {1}{3} 
\int_{- \infty}^{+ \infty} d \xi g(\xi) \xi n_F(\xi)
\label{562r}
\end{equation}
realizing that the two sides of eq.(\ref{560r}) differ in a few per cent.  

In Fig.6 we report the calculated conductivity spectra for $x=0.3$ at 
different temperatures. 
In the large polaron ferromagnetic phase, $\sigma$ is given by 
eq.(\ref{68r}), 
in the small polaron paramagnetic phase by eq.(\ref{60ar}), in the regime of 
coexisting phases by the two preceding conductivities by means of the 
eq.(\ref{14r}). 
With rising temperature, a transfer of spectral weight from low to high 
energies takes place. 
The crossover energy can be estimated about $\omega_c=9 \omega_0$.
When the temperature increases, the large polaron Drude term makes smaller and
at $T=T_c$  the optical response is characterized by the small polaron
absorption band peaked approximatively around $2 g^2 \omega_0$.
Thus changes in the quasi-particle excitations of the system are traceable in
the optical response that, for $T>T_c$, show the opening of an optical gap.
At low temperatures the conductivity shows a clear Drude peak below $\omega_0$
and a quiet asymmetric absorption band with a long tail above its peak 
position at higher energies.
These two types of optical response are due, respectively, to a coherent 
motion and an incoherent absorption band of large polarons.     
In the inset of Fig.6 the Drude peak is shown . As $T$ decreases, the coherent 
response increases continuously.
We stress that the results shown in this figure are all consistent with  
experimental spectra. \cite{kim1,kim2}

In Fig.7 at $T=0.1 T_c$ the optical conductivity is decomposed into two 
components: band (dotted line) and incoherent conductivity (dashed line). 
It is worth-while to notice that the band conductivity provides also the main 
contribution to the polaronic absorption band.
In the inset of Fig.7 the Drude term is fully shown. At low temperatures it can
be obtained by means of the equation

\begin{equation}
\Re \sigma_{Drude}(\omega)=A \frac{\Gamma(\mu)}{\Gamma^2(\mu)+\omega^2} 
\label{600s}
\end{equation}  
where $A$ is a proportionality constant and $\Gamma(\mu)$ is the scattering 
rate of the preceding section calculated at $\mu$.
This inset emphasizes the spectral weight of the absorption band too.

In order to be more quantitative in the comparison with experimental data, we
have evaluated the effective carrier number below the crossover energy 
$\omega_c$

\begin{equation}
N_{eff}(\omega)=\frac{2m_e}{\pi e^2 N_{Mn}} \int_0^{\omega_c}
d\omega \Re \sigma_{\alpha,\alpha}(\omega)    
\label{170ar}
\end{equation}
that is proportional to the spectral weight below this cutoff energy 
(see Fig.8a).
In eq.(\ref{170ar}) $m_e$ is the free electron mass, $e$ is the electron 
charge and $N_{Mn}$ is the number of $Mn$ atoms per unit volume. 
As $T$ increases, $N_{eff}$ decreases assuming a nearly constant value for 
$T>T_c$. 
We stress that the calculated quantity (circles) agrees with the 
experimental effective carrier number deduced by ref.\cite{kim1,kim2} and 
reported in the same figure (diamonds).
We note that the agreement gets worse for $T>T_c$. Indeed, within this 
temperature range, in addition to the small polaron band, other terms, such as
interband transitions between the exchange-split conduction bands \cite{okim}, 
could contribute to the optical absorption.    

We have also estimated the quantity $<\omega>$ (the first moment)

\begin{equation}
<\omega>=\frac{ \int_0^{\infty}d\omega \omega \Re \sigma_{\alpha,\alpha}
(\omega)    }
{\int_0^{\infty}d\omega \Re \sigma_{\alpha,\alpha}(\omega)    }.
\end{equation}
as a function of the temperature (see Fig.8b).
It measures the characteristic energy of the polaron excitations.\cite{calvani}
Thus $<\omega>$ assumes small values in the large polaron phase at low 
temperatures and increases gradually in the phase separation regime tending 
to the value $g^2 \omega_0$ in the small polaron phase at high temperatures.

Finally in Fig.9 we plot the resistivity $\rho=1/\sigma$ , where 
$\sigma$ results from the conductivity $\sigma(\omega)$ in the limit 
$\omega \rightarrow 0$.
We obtain a further confirm of two different behaviors: metallic 
($d\rho/dT>0$) and insulating ($d\rho/dT<0$). In logarithmic scale the 
behavior of activated resistivity \cite{schiffer} for $T>T_c$ is emphasized.

\section{Summary and conclusions} 

We have discussed spectral and optical properties mainly for $x=0.3$ and as 
functions of the temperature. The polaron formation and the small-to-large 
polaron crossover through the phase separation regime turn out to play a 
crucial role in understanding the properties of manganites.  

It has been stressed that the phase separation regime induces a robust
pseudogap in the excitation spectrum of the system. Furthermore the pseudogap
features show a direct relationship with the metal-insulator transition and 
with the crossover from the coherent large polaron dynamics to the incoherent 
small polaron one. Our results turn out compatible with the spectral 
weights found experimentally. \cite{dessau1,dessau2}

With regard to the infrared absorption, we have observed that, with decreasing 
$T$, our optical spectra are characterized by a transfer of spectral weight 
from high to low energies filling up the low frequency optical gap present
in the high-temperature phase.

In the metallic ferromagnetic phase, at low temperatures, the system shows two 
types of optical response: a narrow Drude term and a broad absorption band 
due respectively to the coherent and incoherent motion of large polarons. 
These results obtained at low $T$ are consistent with experimental 
conductivity spectra.\cite{kim1,kim2,okim} 
The agreement improves when a temperature-independent part is extracted  from 
the experimental optical spectra. \cite{okim,okim1} 
This part is considered to be due to the ``background'' interband transitions
between the $O$ $2p$ and $Mn$ $3d$ band.
The reduced optical conductivity obtained by subtracting the 
temperature-independent term \cite{okim} points out that the broad band at low
$T$ is very similar to that shown in this paper.  
We also stress that orbital degrees of freedom could be responsible of the 
effect of enhancing the incoherent absorption. \cite{kilian,horsch}
On the other hand, we observe that electronic correlations can change the 
spectral properties but are somehow obscured in the optical conductivity 
within the ferromagnetic metallic phase. \cite{imai}  

At high temperatures, the infrared absorption is due to the incoherent
small polaron motion. Spectroscopic evidences of the small polaron formation
have been reported in the high-$T$ paramagnetic phase. \cite{yoon,machida} 
However for frequencies higher than the peak of small polaron absorption band,
the effects due to the exchange-split bands and to the local Coulomb repulsion
cannot be neglected. \cite{held}    
Indeed the experimental data can be fitted reasonably well with the small 
polaron band below $0.8$ $eV$. \cite{kim1}
Thus we can consider our results meaningful for frequencies up to the 
absorption peak.

\section*{Acknowledgments}
Valuable discussions with V. Marigliano Ramaglia and F. Ventriglia are
gratefully acknowledged. 
We also thank E. Piegari for a critical reading of the manuscript.
 
\section*{Appendix}

In this appendix we report the calculations leading to the expressions of the 
optical conductivity 
\begin{equation}
\Re \sigma_{\alpha,\alpha}(\omega)=
\Re \sigma^{(band)}_{\alpha,\alpha}(\omega)+
\Re \sigma^{(incoh)}_{\alpha,\alpha}(\omega).
\label{130pr}
\end{equation}
This decomposition originates from the current-current correlation function 
that in Matsubara frequencies can be written as

\begin{equation}
\Pi_{\alpha,\alpha}(i \omega_n)=
\Pi^{(1)}_{\alpha,\alpha}(i \omega_n)+
\Pi^{(2)}_{\alpha,\alpha}(i \omega_n)
\label{120r}
\end{equation}
where the first term is

\begin{equation}
\Pi^{(1)}_{\alpha,\alpha}(i \omega_n)=
4 e^2 t^2 e^{ -2 S_T} 
\left\langle \left( \frac{S_0+1/2}{2 S+1} \right) \right\rangle^2
\left( \frac{1}{N} \right) \sum_{{\bf k}} \sin^2(k_{\alpha})
\int_0 ^{\beta} e^{i \omega_n \tau}
 \tilde{{\cal G}} ({\bf k},-\tau)   
 \tilde{{\cal G}} ({\bf k},\tau)   
\label{121r}
\end{equation}
and the second one reads 

\begin{equation}
\Pi^{(2)}_{\alpha,\alpha}(i \omega_n)=
\left( \frac {2 e^2 } {Z} \right)  
\left\langle \left( \frac{S_0+1/2}{2 S+1} \right) \right\rangle^2
\left( \frac{1}{N^2} \right) \sum_{{\bf k},{\bf k_1}} 
\int_0 ^{\beta} e^{i \omega_n \tau}
\tilde{{\cal G}} ({\bf k},-\tau)
\tilde{{\cal G}} ({\bf k}_1,\tau)
f_1(\tau)     
\label{123r}
\end{equation}
with $f_1(\tau)$ given by eq.(\ref{35r}).
Adopting the Lehmann representation, the spectral function
$\tilde{A} ({\bf k},\omega) $ can be introduced, therefore, making the 
analytic continuation $i \omega_n \rightarrow \omega +i\delta$,  the band 
conductivity becomes 

\begin{eqnarray}
\Re \sigma^{(band)}_{\alpha,\alpha}(\omega)= && 
- \frac{ \Im \Pi^{(1)ret}_{\alpha,\alpha}(\omega)} {\omega}=
\nonumber \\
&& \left( \frac{ 4 e^2 t^2}{\omega} \right) e^{ -2 S_T} 
\left\langle \left( \frac{S_0+1/2}{2 S+1} \right) \right\rangle^2
\left( \frac{1}{N} \right) \sum_{{\bf k}} \sin^2(k_{\alpha})
\tilde{D}( {\bf k} )  
\label{124r}
\end{eqnarray}
where $\tilde{D}({\bf k})=D( {\bf k}, {\bf k} ) $ with 

\begin{equation}
D( {\bf k}, {\bf k}_1 )= 
\int_{- \infty} ^{+ \infty} \frac {d \omega^{\prime}} {4 \pi} 
\tilde{A} ({\bf k},\omega^{\prime})   
\tilde{A} ({\bf k}_1,\omega^{\prime}-\omega)
[ n_F ( \omega^{\prime}-\omega ) -  n_F ( \omega^{\prime} )].  
\label{125r}
\end{equation}
We also derive the latter term of the conductivity

\begin{eqnarray}
\Re \sigma^{(incoh)}_{\alpha,\alpha} && (\omega) =   
- \frac{ \Im \Pi^{(2)ret}_{\alpha,\alpha}(\omega) }{\omega}=
\nonumber \\
&& \left( \frac{ 2 e^2 t^2}{\omega} \right) e^{ -2 S_T} 
\left\langle \left( \frac{S_0+1/2}{2 S+1} \right)^2 \right\rangle
\left( \frac{1}{N^2} \right) \sum_{{\bf k},{\bf k}_1} 
f( {\bf k},{\bf k}_1  )+
\nonumber \\
&& \left( \frac{ 2 e^2 t^2}{\omega} \right) e^{ -2 S_T} 
\left[ I_0(z) \left\langle \left( \frac{S_0+1/2}{2 S+1} \right)^2  
\right\rangle
- \left\langle \left( \frac{S_0+1/2}{2 S+1} \right)  \right\rangle^2  \right]
\left( \frac{1}{N^2} \right) \sum_{{\bf k},{\bf k}_1} 
D( {\bf k}, {\bf k}_1 )
\nonumber \\
\label{126r}
\end{eqnarray}
where

\begin{equation}
f( {\bf k},{\bf k}_1  )=
2\sum_{l=1}^{+ \infty} I_l(z) 
\sinh \left( \frac{\beta \bar{\omega}_0 l}{2} \right)
\left[ 
J_l({\bf k},{\bf k}_1,\omega)+
H_l({\bf k},{\bf k}_1,\omega)
\right]
\label{127r}
\end{equation}
with

\begin{eqnarray}
J_l({\bf k},{\bf k}_1,\omega)= && 
\int_{- \infty} ^{+ \infty} \frac {d \omega^{\prime}} {4 \pi} 
\tilde{A} ({\bf k},\omega^{\prime})   
\tilde{A} ({\bf k}_1,\omega^{\prime}-l\bar{\omega}_0-\omega)
[n_F(\omega^{\prime}-l\bar{\omega}_0-\omega)-
n_F(\omega^{\prime}-l\bar{\omega}_0)]
\times
\nonumber \\
&& \times
\left[ N_0(l \bar{\omega}_0)+n_F(\omega^{\prime}) \right]
\label{128r}
\end{eqnarray}
and

\begin{eqnarray}
H_l({\bf k},{\bf k}_1,\omega)= &&
\int_{- \infty} ^{+ \infty} \frac {d \omega^{\prime}} {4 \pi} 
\tilde{A} ({\bf k},\omega^{\prime})   
\tilde{A} ({\bf k}_1,\omega^{\prime}+l\bar{\omega}_0-\omega)
[n_F(\omega^{\prime}+l\bar{\omega}_0-\omega)-
n_F(\omega^{\prime}+l\bar{\omega}_0)]
\times
\nonumber \\
&& \times
\left[ N_0(l \bar{\omega}_0)+1-n_F(\omega^{\prime}) \right].
\label{129r}
\end{eqnarray}
We observe that $J_l({\bf k},{\bf k}_1,\omega)$ and 
$H_l({\bf k},{\bf k}_1,\omega)$ describe phonon absorption and emission 
processes, respectively. 
Using the convolution product of Lorentzian functions, the band
conductivity becomes

\begin{equation}
\Re \sigma^{(band)}_{\alpha,\alpha}(\omega)= 
\left( \frac{ 4 e^2 t^2}{\omega} \right) e^{ -2 S_T} 
\left\langle \left( \frac{S_0+1/2}{2 S+1} \right) \right\rangle^2
\int_{- \infty}^{+ \infty} d \xi 
 [n_F(\xi-\omega)-n_F(\xi)]
\tilde{C}(\xi,\omega) h(\xi)
\label{130r}
\end{equation} 
where $\tilde{C}(\xi,\omega)=C(\xi,\xi,\omega)$ with   

\begin{equation}
 C(\xi,\xi_1,x)=
\int_{- \infty} ^{+ \infty} \frac {d \omega^{\prime}} {4 \pi} 
\tilde{A} (\xi,\omega^{\prime})   
\tilde{A} (\xi_1,\omega^{\prime}-x)=  
\frac{1}{2}
\frac{ [\Gamma(\xi)+\Gamma(\xi_1)] }
{ [\Gamma(\xi)+\Gamma(\xi_1)]^2/4+(\xi_1-\xi+x)^2 }
\label{131r}
\end{equation} 
and

\begin{equation}
h(\xi)=\left( \frac{1}{N} \right) \sum_{{\bf k}} \sin^2(k_{\alpha})
\delta( \xi - \xi_{{\bf k}} ).
\label{132r}
\end{equation} 
Since $\psi(y)$, the Fourier transform of $ h(\xi) $, is calculated 

\begin{equation}
\psi(y)= \int_{- \infty} ^{+ \infty} d \xi e^{i y \xi} h(\xi)=
\frac { [ J_0( 2 y t_{eff} ) ]^2 J_1 (2 y t_{eff}) } {2 y t_{eff}}
\label{133r}
\end{equation}
we can express $ h(\xi) $ in the following manner

\begin{equation}
h(\xi)= \frac{1}{ 2 \pi t_{eff}}
\int_{- \infty} ^{+ \infty} dy \cos \left( \frac {\xi}{2 t_{eff}} \right)  
\frac { [J_0(y) ]^2 J_1(y) }
{y }.
\label{133br}
\end{equation} 
On the other hand, the latter part of the conductivity reads

\begin{eqnarray}
\Re \sigma^{(incoh)}_{\alpha,\alpha}(\omega)= && 
\left( \frac{ 2 e^2 t^2}{\omega} \right) e^{ -2 S_T} 
\left\langle \left( \frac{S_0+1/2}{2 S+1} \right)^2 \right\rangle
\int_{- \infty} ^{+ \infty} d \xi 
\int_{- \infty} ^{+ \infty} d \xi_1  
g(\xi) g(\xi_1) R(\xi,\xi_1,\omega)+
\nonumber \\
&& \left( \frac{ 2 e^2 t^2}{\omega} \right) e^{ -2 S_T} 
\left[ I_0(z) \left\langle \left( \frac{S_0+1/2}{2 S+1} \right)^2  
\right\rangle
- \left\langle \left( \frac{S_0+1/2}{2 S+1} \right)  \right\rangle^2  \right]
\times
\nonumber \\
&& \times
\int_{- \infty} ^{+ \infty} d \xi 
\int_{- \infty} ^{+ \infty} d \xi_1  
g(\xi) g(\xi_1)  [n_F(\xi-\omega)-n_F(\xi)]
C(\xi,\xi_1,\omega) 
\label{134r}
\end{eqnarray}
where $g(\xi)$ is the density of states (\ref{12r}) and the function
$ R(\xi,\xi_1,\omega) $ is given by 

\begin{equation}
 R(\xi,\xi_1,\omega)=
2\sum_{l=1}^{+ \infty} I_l(z) 
\sinh \left( \frac{\beta \bar{\omega}_0 l}{2} \right)
\left[ 
J_l(\xi,\xi_1,\omega)+
H_l(\xi,\xi_1,\omega)
\right]
\label{135r}
\end{equation}
with

\begin{eqnarray}
J_l( \xi,\xi_1,\omega )=
C( \xi,\xi_1,\omega+l\bar{\omega}_0 ) 
[n_F(\xi-l\bar{\omega}_0-\omega)-
n_F(\xi-l\bar{\omega}_0)]
\left[ N_0(l \bar{\omega}_0)+n_F(\xi) \right]
\label{136r}
\end{eqnarray}
and
 
\begin{eqnarray}
H_l( \xi,\xi_1,\omega )=
C( \xi,\xi_1,\omega-l\bar{\omega}_0 )
[n_F(\xi+l\bar{\omega}_0-\omega)-
n_F(\xi+l\bar{\omega}_0)]
\left[ N_0(l \bar{\omega}_0)+1-n_F(\xi) \right].
\label{137r}
\end{eqnarray} 

In the limit of high temperatures, we concentrate on $f_1(\tau)$ of 
eq.(\ref{123r}) carrying out the following approximation

\begin{equation}
e^{-2 S_T} 
\exp \left\{
z\cosh\left[\bar{\omega}_0\left(\tau-\frac{\beta}{2}\right)\right]
\right\} 
\simeq
\exp (- \bar{z} \tau) 
\exp \left( \frac{\bar{z}}{\beta} \tau^2 \right) 
\end{equation}
where

\begin{equation}
\bar{z}= 2 \bar{s} \simeq 2 g^2 \omega_0.
\end{equation}
This expansion is valid as long as 
$ \left( \frac{ \beta \bar{z}}{12} \right) 
 \left( \frac{ \beta \bar{\omega}_0}{4} \right)^2 <1 $, 
hence $T>0.39 \omega_0 $.
The response of the small polaron excitations is characterized by the 
incoherent behavior, therefore the conductivity becomes 

\begin{eqnarray}
\Re \sigma_{\alpha,\alpha} (\omega)=
\nonumber \\
\left( \frac{e^2 t^2}{\omega} \right)  
\left\langle \left( \frac{S_0+1/2}{2 S+1} \right)^2 \right\rangle
&& \sqrt{ \frac{ \pi \beta}{ \bar{z}}}
\left( \frac{1}{N^2} \right) \sum_{\bf{k},\bf{k}_1}
\int_{- \infty} ^{+ \infty} \frac{d \omega^{\prime}}{2 \pi} 
\int_{- \infty} ^{+ \infty} \frac {d \omega^{\prime \prime}}{2 \pi}  
 \tilde {A} \left( { \bf k}, \omega^{\prime} \right)  
\tilde {A} \left( {\bf k}_1, \omega^{\prime \prime} \right)
\times
\nonumber \\
\times
\left[ 1-n_F ( \omega^{\prime \prime} ) \right] n_F(\omega^{\prime} )
&& 
\left\{ \exp \left[  - \frac{\beta}{4 \bar{z}} 
( \omega +\omega^{\prime} -\omega^{\prime \prime} - \bar{z} )^2  \right] -
\exp \left[  - \frac{\beta}{4 \bar{z}} 
(\omega +\omega^{\prime} -\omega^{\prime \prime} +\bar{z} )^2  \right]
\right\}.
\nonumber \\
\label{50r}
\end{eqnarray}
We consider $ \xi_{{\bf k}} \simeq -g^2 \omega_0 - \mu $ obtaining

\begin{eqnarray}
&& \Re \sigma_{\alpha,\alpha}(\omega)= 
\nonumber \\
&& \left( \frac{e^2 t^2}{\omega} \right)  
\left\langle \left( \frac{S_0+1/2}{2 S+1} \right)^2 \right\rangle
\sqrt{ \frac{ \pi \beta}{ \bar{z}}}
\left\{
\exp \left[  - \frac{\beta}{4 \bar{z}} 
( \omega - \bar{z} )^2  \right] -
\exp \left[  - \frac{\beta}{4 \bar{z}} 
(\omega + \bar{z} )^2  \right]
\right\}
\rho ( 1- \rho )
\label{60r}
\end{eqnarray}
where $\rho$ is the electron concentration.

\section*{Figure captions}
\begin{description}

\item  {F1} 
(a) The phase diagram corresponding to $t=2\omega_0$, $g=2.8$ and 
$\epsilon=0.01t$. $PI$ means Paramagnetic Insulator, $FM$ Ferromagnetic Metal 
and $AFI$ AntiFerromagnetic Insulator.
The areas $PI+FM$ and $AFI+FM$ indicate regions where localized ($PI$ and 
$AFI$) and delocalized ($FM$) phases coexist.
(b) The fractions of volume $(V_1/V)_{ferro}$ and $(V_2/V)_{para}$ at $x=0.3$ 
as a function of the temperature ($T$ is expressed in units of $\omega_0$).

\item  {F2} 
The spectral function (corresponding to  $t=2\omega_0$ and $g=2.8$) at $x=0.3$
as a function of the energy (in units of $\omega_0$) for two different
temperatures: (solid line) $T=0.05$ and $\xi_{\bf{k}}=0$, (dashed line) 
$T=0.47$ ($T$ is expressed in units of $\omega_0$). 
 
\item  {F3} 
The renormalized density of states at $x=0.3$ as a function of the
energy (in units of $\omega_0$) for different temperatures. In the inset the 
renormalized density of states at $\mu$ as a function of the temperature.  
The results are obtained for $t=2\omega_0$ and $g=2.8$.

\item  {F4} 
The renormalized density of states at $T=0.39 \omega_0$
as a function of the energy (in units of $\omega_0$) for different
densities. In the inset the renormalized density of states in a small range 
of energy around the chemical potential $\mu$. 

\item  {F5} 
(a) The scattering rate (corresponding to $t=2\omega_0$ and $g=2.8$) at 
$x=0.3$ as a function of the energy (in units of $\omega_0$) for different 
temperatures.  
(b) The scattering rate (solid line), 1-phonon rate (dotted line), 
spin-fluctuation rate (dashed line) and multi-phonon rate (dotted-dashed line)
at $x=0.3$ and $T=0.65 T_c$ as a function of the energy 
(in units of $\omega_0$).

\item  {F6} 
The conductivity up to 18 $\omega_0$ at different temperatures.
In the inset the conductivity up to 1 $\omega_0$ at different temperatures.
The conductivities are expressed in units of $\frac{e^2c}{m \omega_0}$,
with $c$ hole concentration and $m= \frac {1}{2t}$, and they are obtained for 
$t=2\omega_0$ and $g=2.8$.

\item  {F7} 
The conductivity (solid line), the band conductivity 
(dotted line) and the ``incoherent'' conductivity (dashed line) at $T=0.1 T_c$ 
in units of $\frac{e^2c}{m \omega_0}$, where $c$ is the hole concentration and
$m= \frac {1}{2t}$.
In the inset the conductivity (solid line) and the Drude term (dotted line) at
$T=0.1 T_c$ in the same units of the figure.

\item  {F8}
(a) The effective carrier number (circles) below the crossover energy 
$\omega_c $ compared to the experimental effective carrier number deduced by 
ref.\cite{kim1,kim2} as a function of the temperature.
(b) The first moment (in units of $\omega_0$) as a function of the temperature.
The results are obtained for $t=2\omega_0$ and $g=2.8$, in the case (a)
specifying $\omega_0=50$ $meV$ and the lattice constant $a=0.4$ $nm$.

\item  {F9} 
The resistivity as a function of the temperature (we have used 
$\omega_0=50$ $meV$ and the lattice constant $a=0.4$ $nm$).

\end{description}

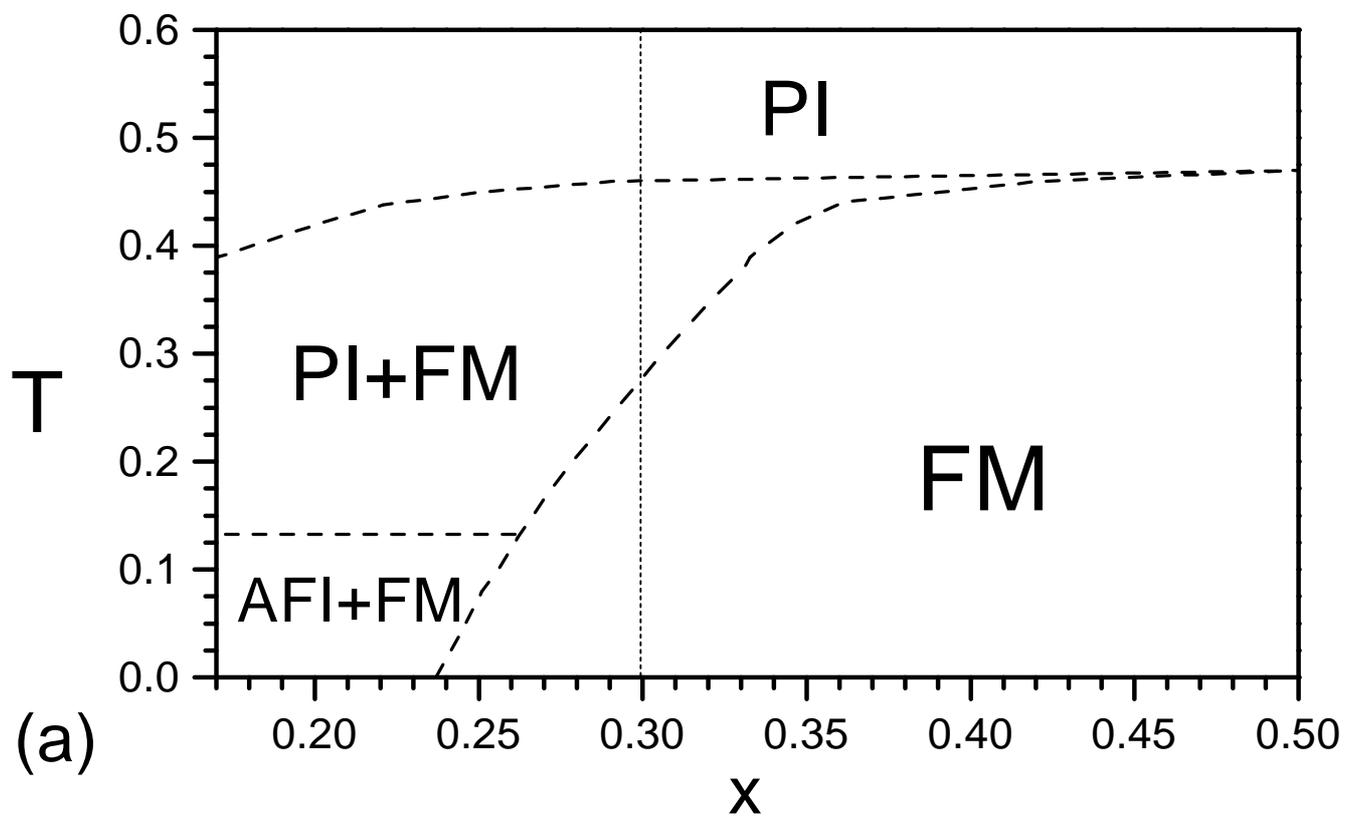
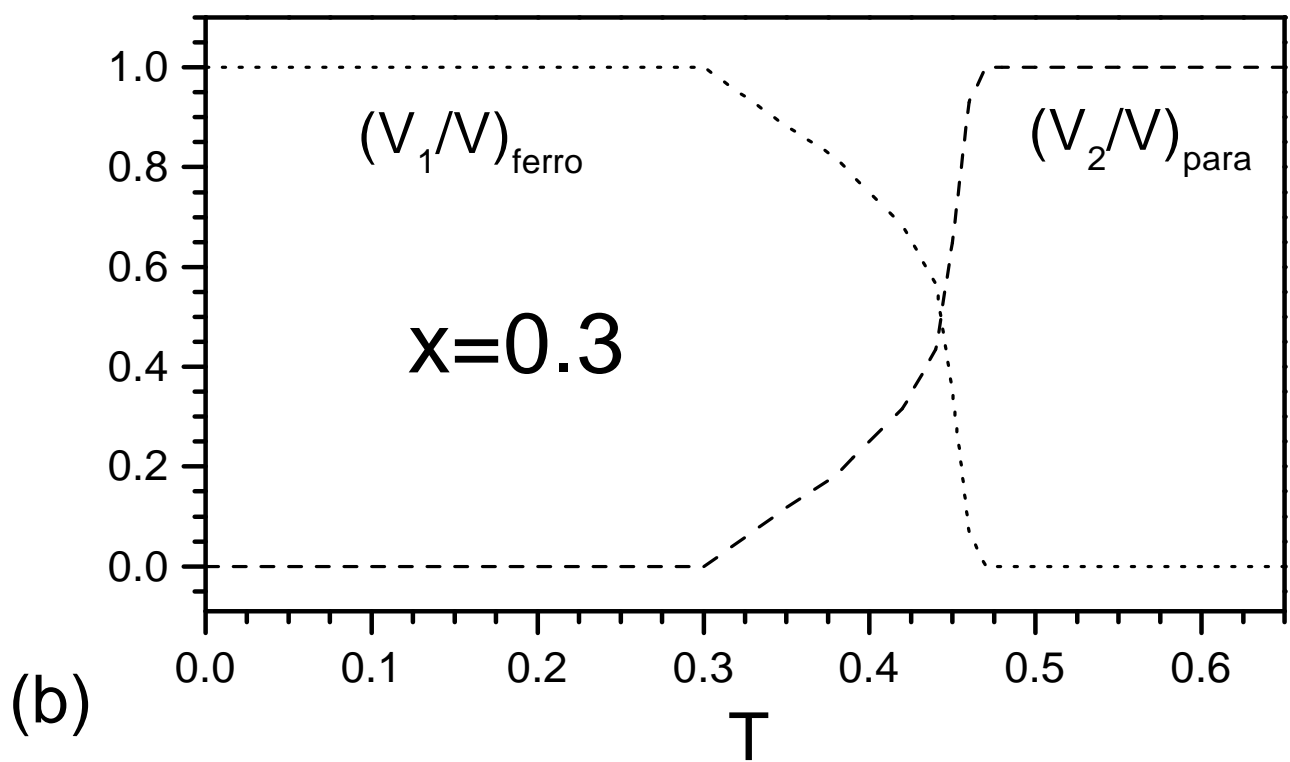
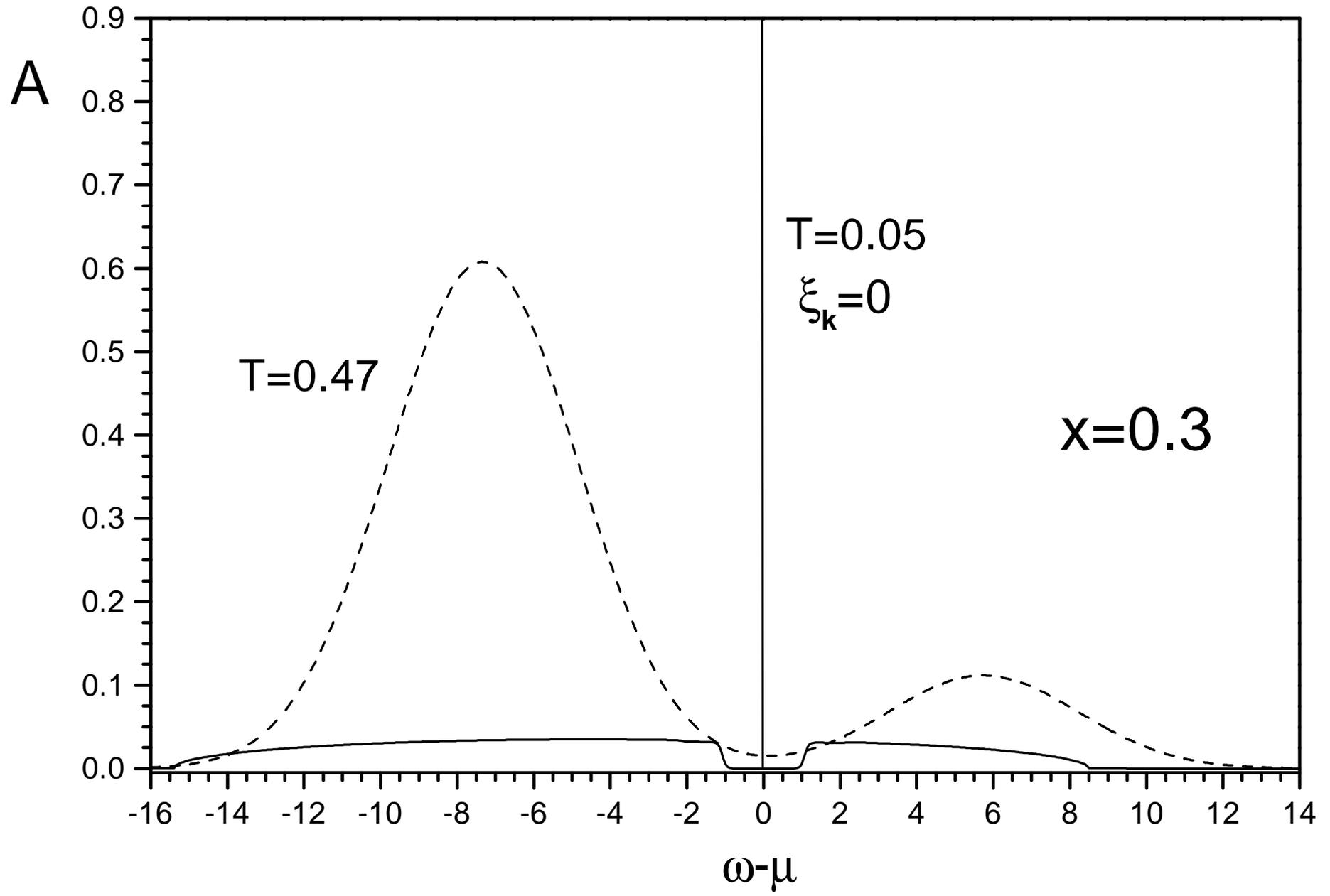
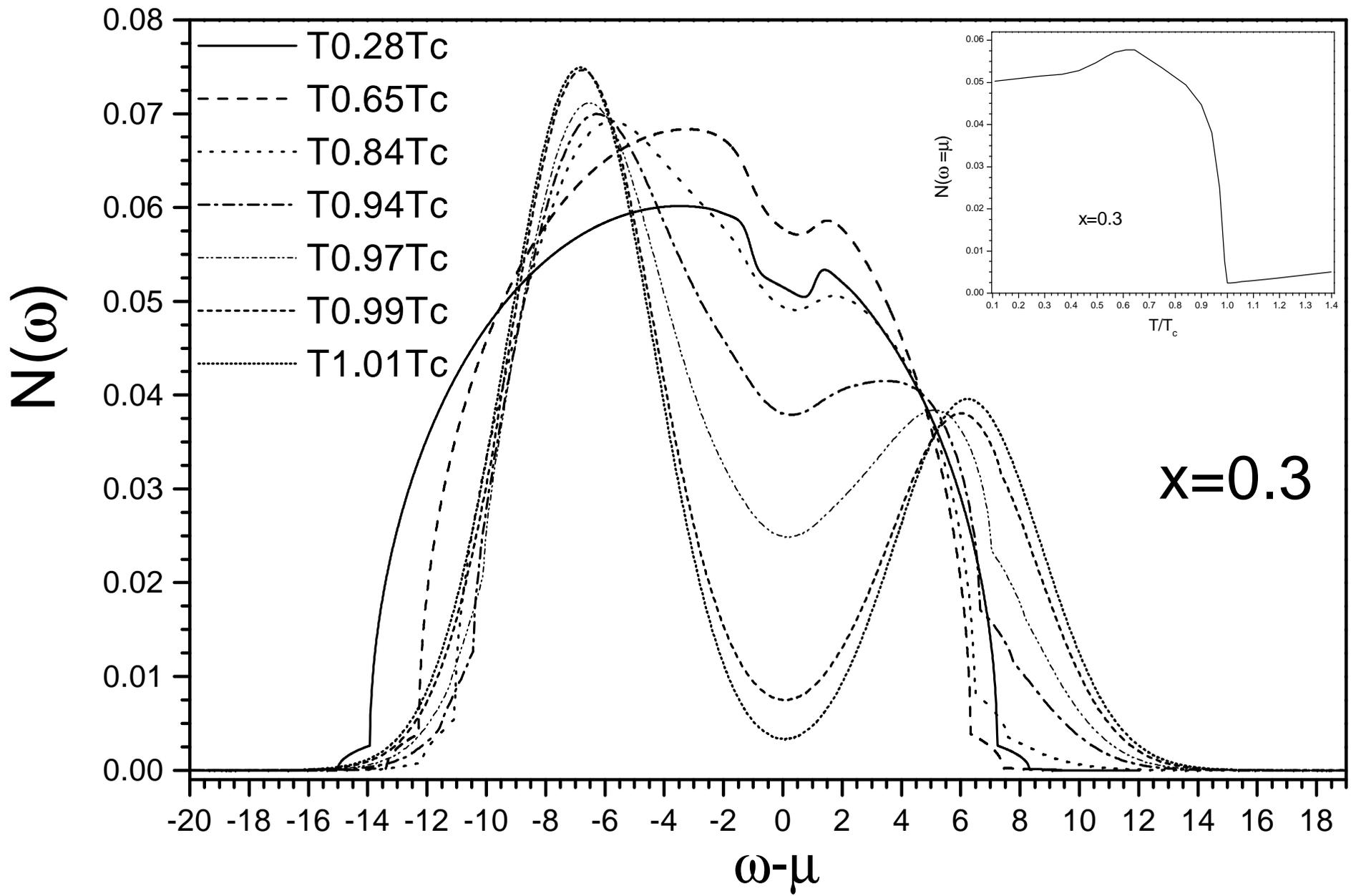
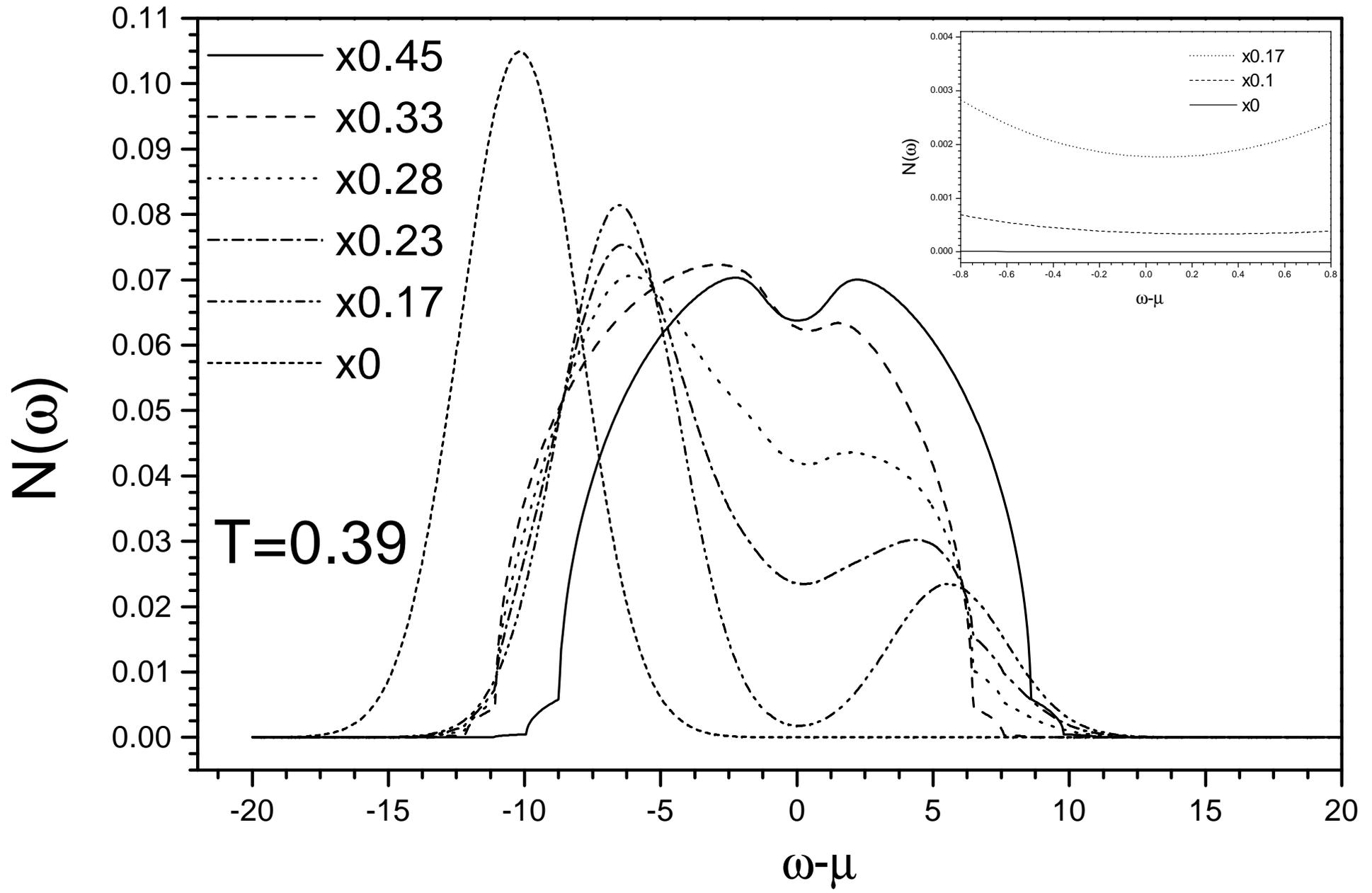
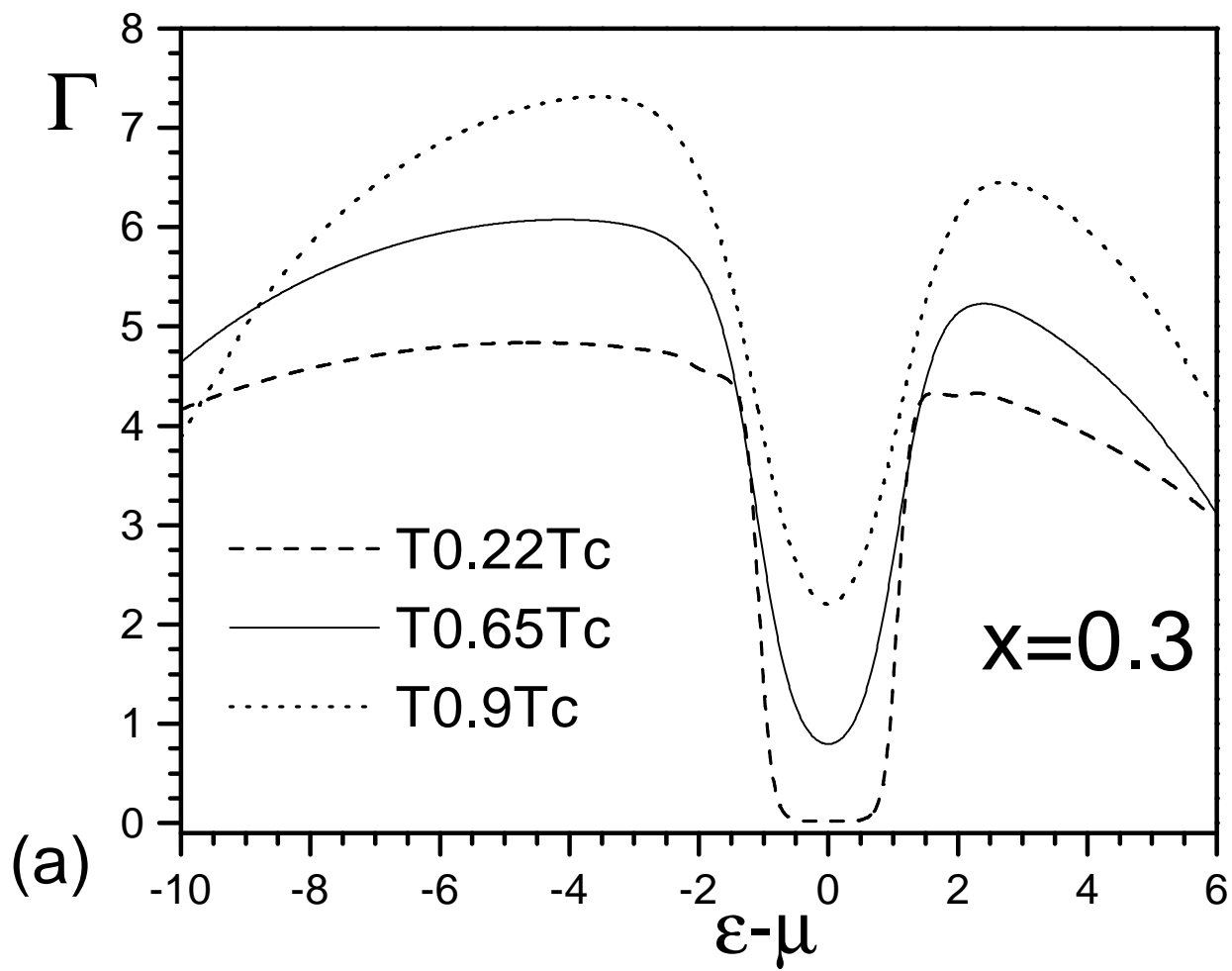
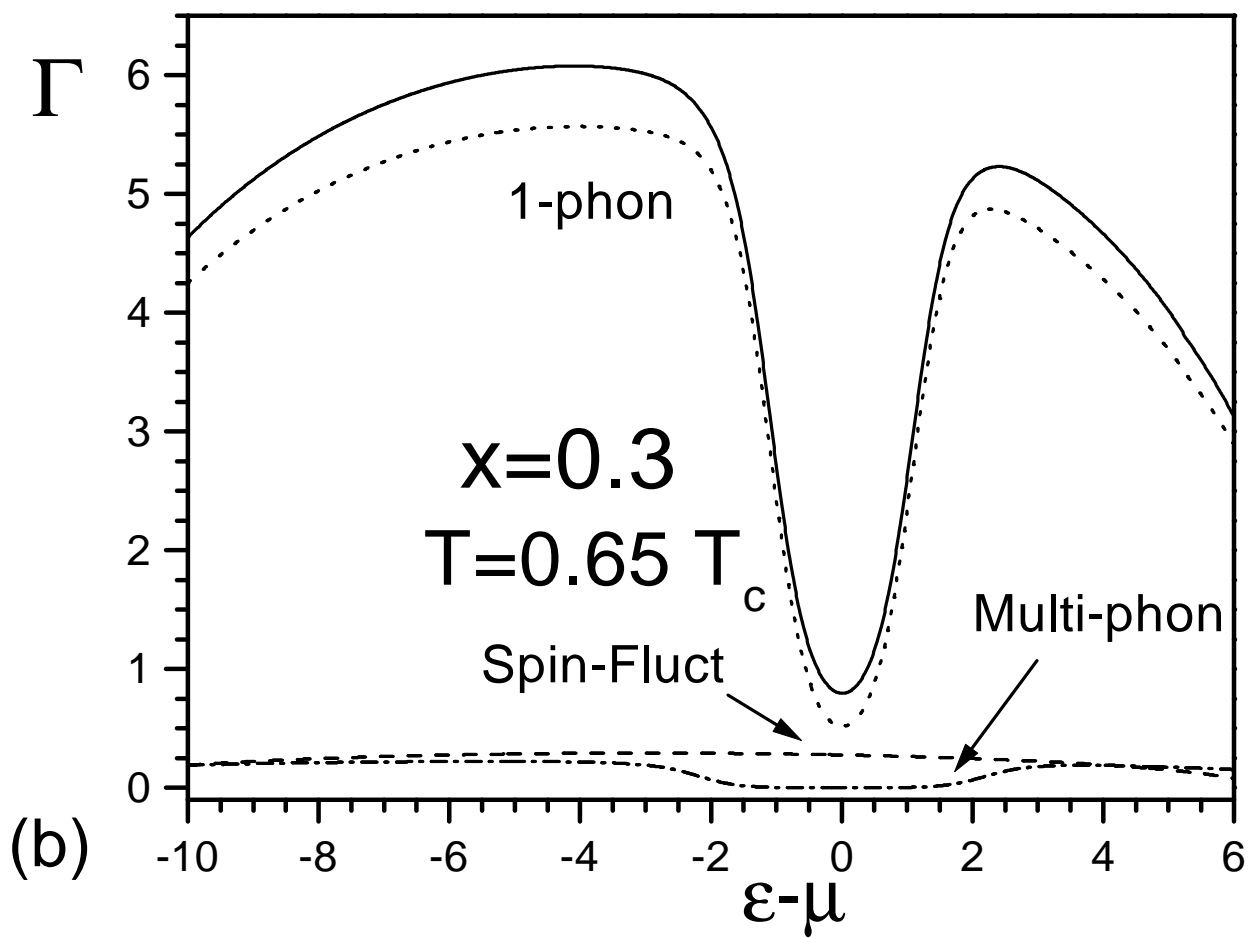
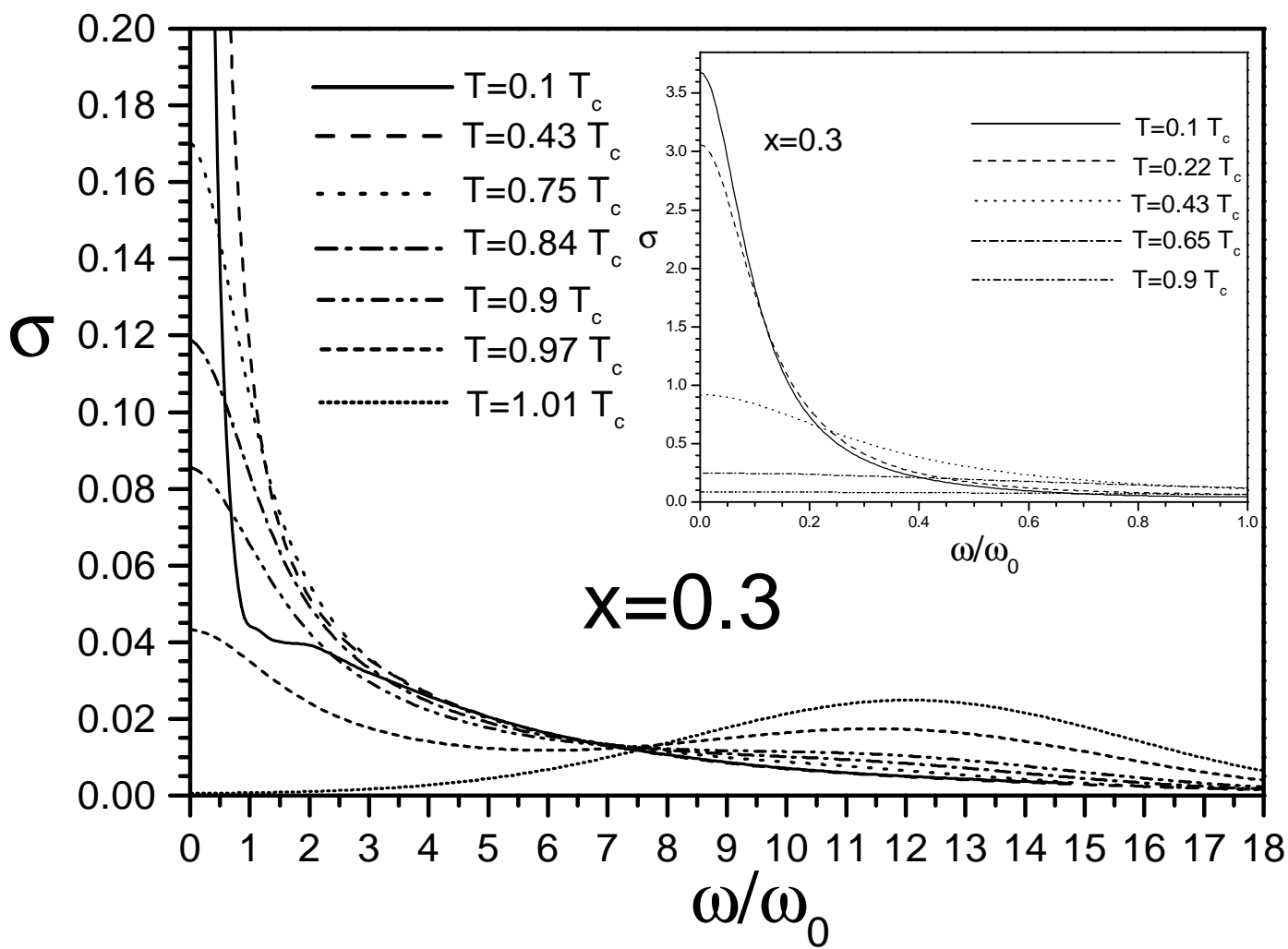
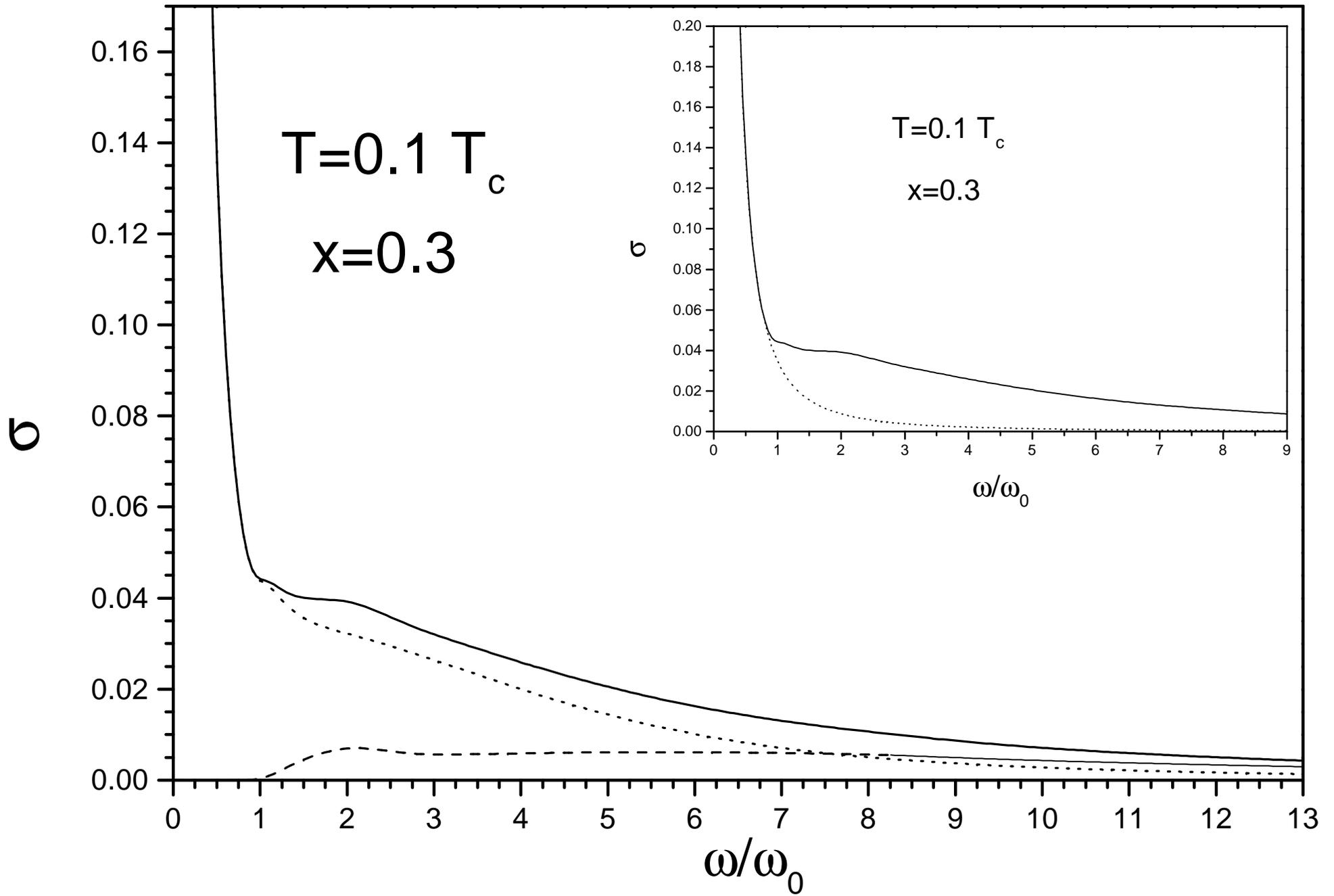
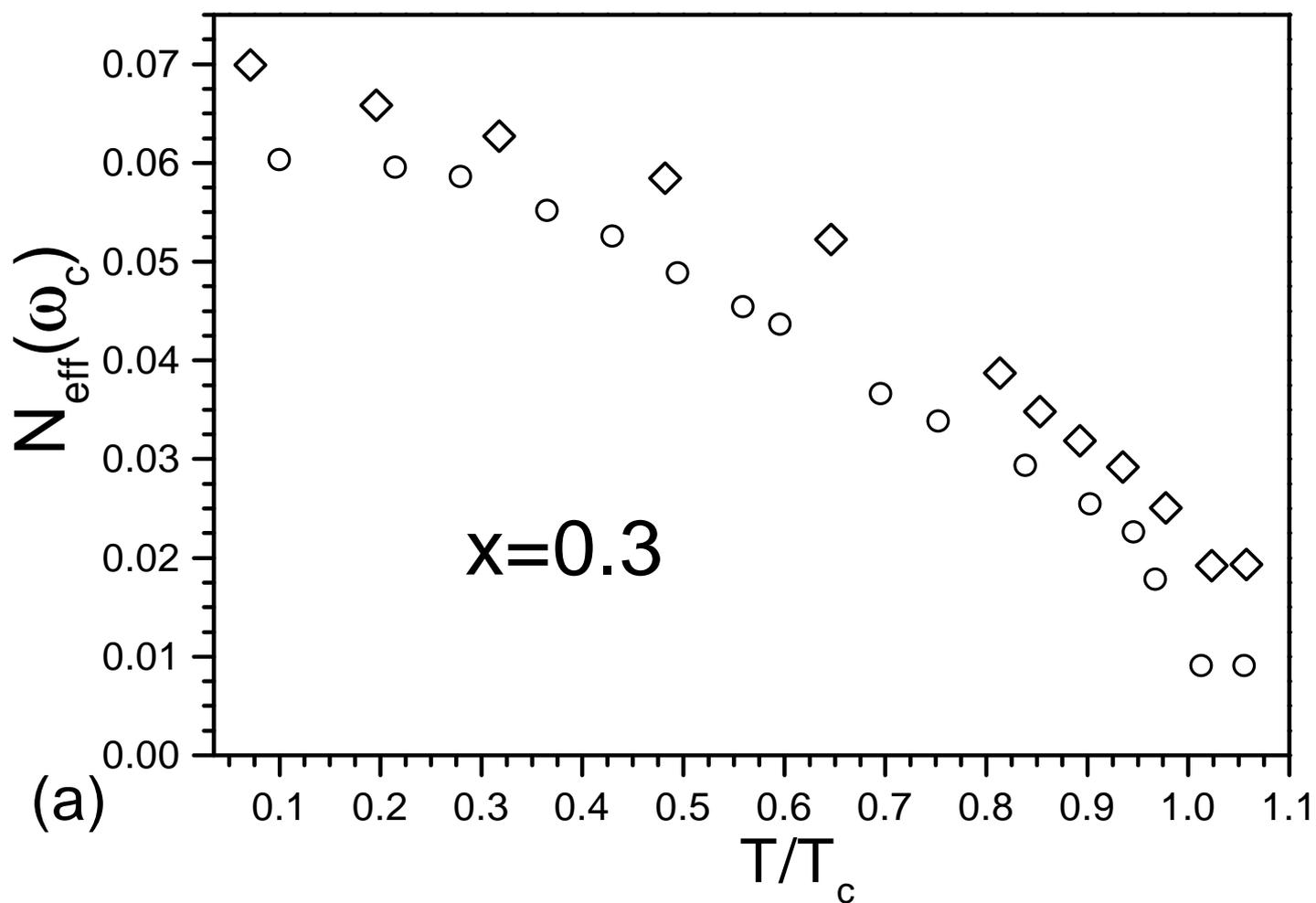
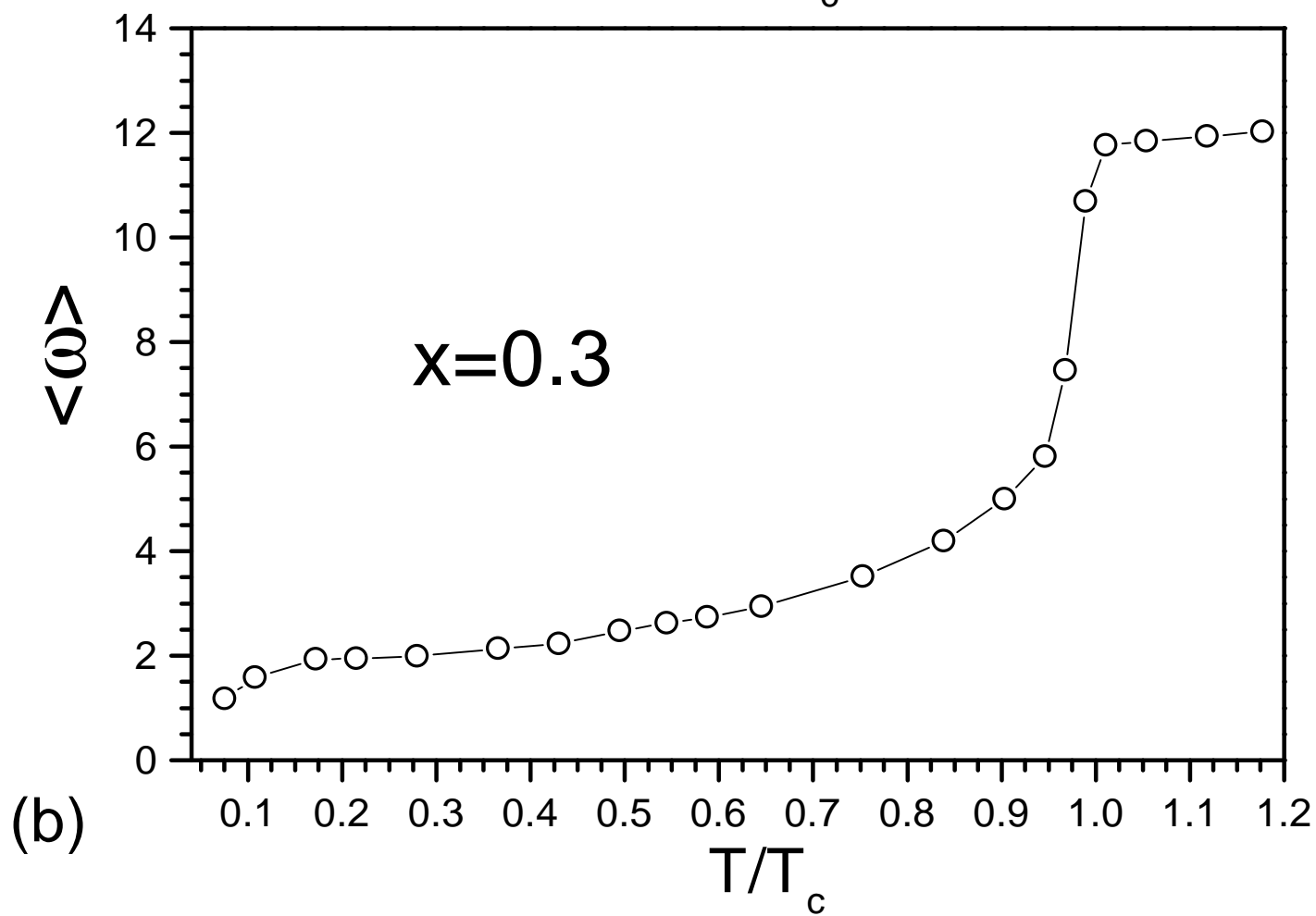
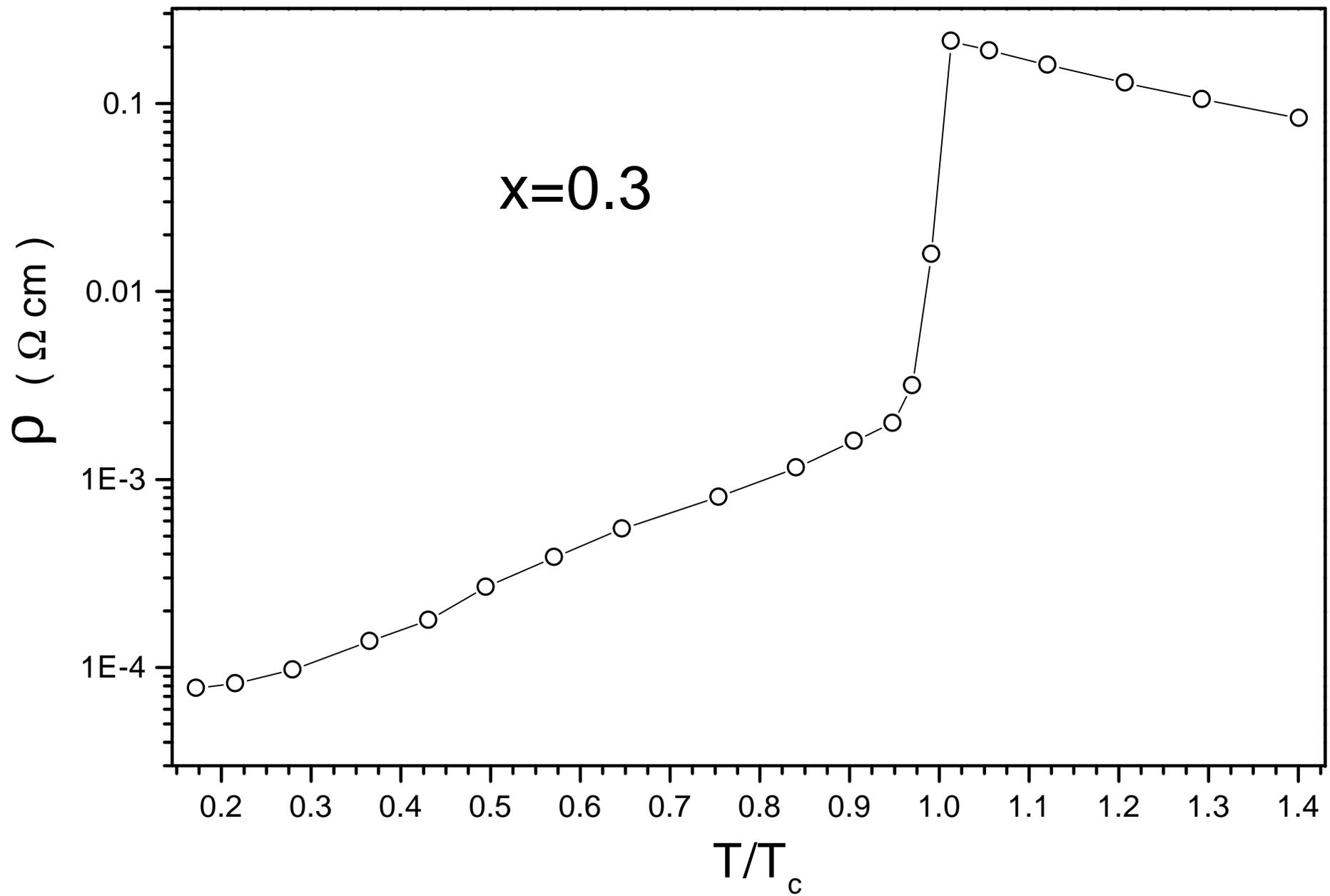
\end{document}